\documentclass[aps,prd,twocolumn,nofootinbib,preprintnumbers,superscriptaddress,showpacs,amsmath,amssymb]{revtex4-1}
\usepackage{ulem} 
\usepackage{booktabs,makecell}
\usepackage{graphicx}
\usepackage{amsmath,amssymb}
\usepackage{amsfonts}
\usepackage{xspace} 
\usepackage[usenames]{color}
\usepackage{dcolumn}
\usepackage{bm}
\usepackage{mathrsfs}
\usepackage[colorlinks=true]{hyperref}
\usepackage[all]{hypcap} 
\usepackage[utf8]{inputenc} 
\usepackage{slashed}
\usepackage{multirow}
\usepackage{rotating}
\usepackage{nicematrix}
\definecolor{orange}{rgb}{1,0.5,0}

\begin{document}
%
%

\title{Hybrid star phenomenology from the properties of the special point}

\author{Christoph Gärtlein} \email{christoph.gartlein@tecnico.ulisboa.pt} 
\affiliation{Centro de Astrof\'{\i}sica e Gravita\c c\~ao  - CENTRA, Departamento de F\'{\i}sica, Instituto Superior T\'ecnico - IST, Universidade de Lisboa - UL, Av. Rovisco Pais 1, 1049-001 Lisboa, Portugal}
\affiliation{CFisUC, Department of Physics, University of Coimbra, 3004-516 Coimbra, Portugal}
\affiliation{Institute of Theoretical Physics, University of Wroclaw, 50-204 Wroclaw, Poland}

\author{Oleksii Ivanytskyi} \email{oleksii.ivanytskyi@uwr.edu.pl}  
\affiliation{Incubator of Scientific Excellence---Centre for Simulations of Superdense Fluids, University of Wrocław, 50-204, Wroclaw, Poland}

\author{Violetta Sagun} 
\email{violetta.sagun@uc.pt}
\affiliation{CFisUC, Department of Physics, University of Coimbra, 3004-516 Coimbra, Portugal}

\author{David Blaschke}
\email{david.blaschke@uwr.edu.pl} 
\affiliation{Institute of Theoretical Physics, University of Wroclaw, 50-204 Wroclaw, Poland}
\affiliation{Helmholtz-Zentrum Dresden-Rossendorf (HZDR), Bautzner Landstrasse 400, 01328 Dresden, Germany}
\affiliation{Center for Advanced Systems Understanding (CASUS), Untermarkt 20, 02826 G\"orlitz, Germany}

\date{\today}
\begin{abstract}

We study the properties of hybrid stars containing a color superconducting quark matter phase in their cores, which is described by the chirally symmetric formulation of the confining relativistic density functional approach. It is shown that depending on the dimensionless vector and diquark couplings of quark matter, the characteristics of the deconfinement phase transition are varied, allowing us to study the relation between those characteristics and mass-radius relations of hybrid stars. Moreover, we show that the quark matter equation of state (EoS) can be nicely fitted by the Alford-Braby-Paris-Reddy model that gives a simple functional dependence between the most important parameters of the EoS and microscopic parameters of the initial Lagrangian. Based on it, we analyze the special points of the mass-radius diagram in which several mass-radius curves intersect.
Using the found empirical relation between the mass of the special point, the maximum mass of the mass-radius curve, and the onset mass of quark deconfinement, we constrain the range of values of the vector and diquark couplings of the quark matter model.
With this constraint, we construct a family of mass-radius curves, which allow us to describe the black widow
pulsar PSR J0952-0607 with a mass of $2.35\pm 0.17~\rm M_\odot$ as a hybrid star with a color superconducting quark matter core.

\end{abstract}
\maketitle
\section{Introduction}
\label{intro}

The equation of state (EoS) of cold, dense strongly interacting matter is presently not accessible by ab initio simulations of quantum chromodynamics (QCD) on the lattice, due to the sign problem in evaluating the partition function.
Therefore, effective approaches to the EoS need to be developed that can capture basic features of QCD, like symmetries of the Lagrangian and their dynamical symmetry-breaking patterns, in a quantifiable way \cite{Fukushima:2010bq}. 
A central role in developing an effective low-energy QCD Lagrangian has been played by the Nambu--Jona-Lasinio (NJL) model \cite{Klevansky:1992qe,Buballa:2003qv}.
It is able to address, in particular, the dynamical chiral symmetry breaking accompanied by the formation of the pion as a pseudo-Goldstone boson and color superconductivity in cold, dense quark matter due to diquark condensation \cite{Alford:1998mk} as an inevitable consequence of the Cooper theorem \cite{Cooper:1956zz}.
The lack of dynamical quark confinement in the NJL model could be partially compensated by assuming a confining bag pressure. 
An effective model for the quark matter EoS that captures the aspects of confinement and color superconductivity, as well as perturbative QCD corrections, is the model by Alford, Braby, Paris, and Reddy (ABPR) \cite{Alford:2004pf} which found a broad application in studying the phenomenology of NSs (see, e.g., \cite{Ozel:2010bz, Blaschke:2021poc, Blaschke:2022egm}).
This model, however, is not defined by an effective Lagrangian and therefore plays only a limited role in elucidating low-energy QCD. 
While NJL models require superimposing a phenomenological confining parameter like a bag constant, we want to base our study on the recently developed confining relativistic density functional (RDF) approach \cite{Kaltenborn:2017hus} in its recent form with a chiral symmetry and color superconductivity \cite{Ivanytskyi:2022oxv}.
We also present a simple and easy-to-use analytical parameterization of the quark matter EoS given in the ABPR form with the parameters directly related to the coupling constants of the Lagrangian of the RDF approach.
This parameterization allows us to discuss relation between the non-perturbative RDF approach and the asymptotics of perturbative QCD within the strategy suggested in Ref. \cite{Blaschke:2022egm}.

Because of the unique mapping between the EoS and neutron star (NS) structure by the Tolman-Oppenheimer-Volkoff (TOV) equations, the measurement of masses and radii of NSs can provide direct constraints for the cold dense matter EoS \cite{Baym:2017whm}, like the lattice QCD simulations benchmark EoS models at finite temperature.
The measurement of bulk properties, however, may leave some degeneracy in the description since the hybrid stars with deconfined quark matter interior may sometimes masquerade themselves as NSs \cite{Alford:2004pf}. 

Despite the fact that macroscopic properties of NSs depend on the strongly interacting matter EoS, special points (SPs) of the NS mass-radius diagram exhibit weak sensitivity to details of such an EoS.
These SPs are narrow ranges of the intersection of multiple mass-radius curves corresponding to different EoS of strongly interacting matter \cite{Yudin:2014mla}.
They exclusively appear only within the scenario of hybrid quark-hadron EoS and are almost insensitive to both properties of hadron matter and details of the quark-to-hadron phase transition \cite{Cierniak:2020eyh,Blaschke:2020vuy,Cierniak:2021knt,Cierniak:2021vlf}.
This makes SPs attractive in the context of extracting model-independent information about the properties of quark matter.
More specifically, we show how the physical parameters of the mass-radius curves corresponding to a given SP are related to each other via a dependence controlled by the NS mass in this SP.
This dependence can be extracted from the mass-radius curves intersecting in their SP and does not require any additional observational input information. 
This allows us to utilize the present observational constraints on the maximum mass of NS in order to restrict the parameter range for the microscopic quark matter model.

The paper is organized as follows. In the next section, we present the RDF approach to model quark matter and show how it can be fitted by the ABPR model. 
In Section  \ref{sec3} we study the properties of hybrid stars and the SPs.
Finally, in Section \ref{concl} we present the main conclusions of our work and discuss its phenomenological outcomes. 
In this study, we adopt the negative metric signature $(+---)$ and geometrical units ($\hbar=c=G=1$).

\section{EoS of quark matter}
\label{sec1}

The main aspects of quark matter, which should be accounted for within a realistic approach, correspond to quark confinement at low densities, dynamical restoration of chiral symmetry at intermediate densities, and consistency with the vacuum phenomenology of QCD.
Formation of the color superconducting state and vector repulsion in dense quark matter, which can be motivated by various gluon exchange channels, are also important for such approaches in the context of phenomenology of NSs \cite{Baym:2017whm}. 
These issues have been recently addressed within a chirally symmetric formulation of the confining RDF approach for color superconducting quark matter \cite{Ivanytskyi:2021dgq,Ivanytskyi:2022oxv}. 
The corresponding EoS can be further constrained by the additional requirement of its conformal behavior at asymptotically high densities. 
This sets constraints on the strength of quark-quark interaction in the vector and diquark channels at high densities.
Recently such constraints were incorporated to a version of the confi\-ning RDF approach for asymptotically conformal quark matter with color superconductivity and vector repulsion \cite{Ivanytskyi:2022bjc}.
In this section, we outline its main aspects.
We also parameterize this microscopic approach by the ABPR parameterization of quark EoS, which by construction respects conformal limit.
The consideration is limited to the two-flavor case, while its generalization to three quark flavors was suggested in Ref. \cite{Blaschke:2022knl}. 

\subsection{Confining RDF approach}
\label{sec2_1}

Quark fields are represented by the flavor spinor $q^T=(u,d)$ entering the Lagrangian
\begin{eqnarray}
\label{I}
\mathcal{L}&=&\overline{q}(i\slashed\partial- m)q+\mathcal{L}_V+\mathcal{L}_D-\mathcal{U},
\end{eqnarray}
where $m$ is the current mass for simplicity chosen to be the same for two light quark flavors. Repulsive vector and attractive diquark paring channels
are responsible for stiffening the quark EoS at high densities and formation of its two-flavor color superconducting (2SC) phase, respectively.
The interaction strength in these channels is controlled by the couplings $G_V$ and $G_D$ entering the model through
\begin{eqnarray}
\label{II}
\mathcal{L}_V&=&-G_V(\overline{q}\gamma_\mu q)^2+\Theta_V,\\
\label{III}
\mathcal{L}_D&=&G_D
(\overline{q}i\gamma_5\tau_2\lambda_A q^c)(\overline{q}^ci\gamma_5\tau_2\lambda_A q)-\Theta_D.
\end{eqnarray}
In the case of $\mathcal{L}_D$ summation over the dummy color index is performed for $A=2,5,7$ with $\lambda_A/2$ being generators of SU(3) color group, while charge conjugation is performed as $q^c=i\gamma_2\gamma_0\overline{q}^T$. 
Recovering the conformal limit within this approach is provided by the behavior of the density-dependent vector and diquark couplings, which is motivated by non-perturbative gluon exchange of QCD in the Landau gauge \cite{Song:2019qoh}. 
This approach accounts for the non-perturbative nature of interaction between quarks in the density range typical for NSs by the non-perturbative gluon mass $M_g$, generated by spontaneous symmetry breaking \cite{Cornwall:1981zr}.
The functional dependences of the vector and diquark couplings on the quark number density $\langle q^+ q\rangle$ and diquark condensate $\langle\overline{q}^ci\tau_2\gamma_5\lambda_2q\rangle$ are given by
\begin{eqnarray}
\label{IV}
G_V&=&\frac{G_{V0}}{1+\frac{8}{9M_g^2}\left(\frac{\pi^2 \langle q^+q\rangle}{2}\right)^{2/3}}
\end{eqnarray}
and 
\begin{eqnarray}
\label{V}
G_D&=&\frac{G_{D0}}{1+\frac{8}{9M_g^2}\left(\frac{\pi^2 |\langle\overline{q}^ci\tau_2\gamma_5\lambda_2q\rangle|}{2}\right)^{2/3}},
\end{eqnarray}
respectively, where $G_{V0}$ and $G_{D0}$ are vacuum values of these couplings.
In Ref. \cite{Ivanytskyi:2022bjc} the value $M_g=600$ MeV was shown to be consistent with the Shifman-Vainshtein-Zakharov expansion of the two-point current correlation functions within massive gauge invariant QCD \cite{Graziani:1984cs} and simultaneously providing good agreement with observational constraints on masses, radii and tidal deformability of NSs.
Further analysis is performed for this value of the non-perturbative gluon mass.

The thermodynamic consistency of this approach is provided by the properly defined density-dependent rearrangement terms 
\begin{eqnarray}
\label{VI}
\Theta_V&=&\int\limits_0^{\langle q^+q\rangle} dn~n^2~\frac{\partial G_V(n)}{\partial n},\\
\label{VII}
\Theta_D&=&\int\limits_0^{|\langle\overline{q}^ci\tau_2\gamma_5\lambda_2 q\rangle|} dn~n^2~\frac{\partial G_D(n)}{\partial n}.
\end{eqnarray}
In the case of constant vector and diquark couplings ($M_g\rightarrow\infty$) these $\Theta_V$ and $\Theta_D$ vanish.

Proper chiral dynamics of the model is induced by the attractive interaction in scalar and pseudoscalar channels.
Chiral symmetry of the corresponding potential $\mathcal{U}$ is provided by choosing its argument in a chirally symmetric form. 
In this work, we consider the form motivated by the String Flip model \cite{Horowitz:1985tx,Ropke:1986qs} suggested in Refs. \cite{Ivanytskyi:2021dgq,Ivanytskyi:2022oxv}
\begin{eqnarray}
\label{VIII}
\mathcal{U}=D_0\left[(1+\alpha)\langle \overline{q}q\rangle_0^2
-(\overline{q}q)^2-(\overline{q}i\gamma_5\vec\tau q)^2\right]^{\frac{1}{3}}.
\end{eqnarray}
Here $\langle \overline{q}q\rangle_0$ is the vacuum value of chiral condensate, whereas the constant $D_0$ and $\alpha\ge0$ will be discussed below. Strong non-linearity of this potential can be overcome by expanding it around the mean-field solutions $\langle \overline{q}q\rangle$ and $\langle \overline{q}i\gamma_5\vec\tau q\rangle=0$. We limit ourselves to the second order leading to
\begin{eqnarray}
\mathcal{U}^{(2)}&\simeq&\mathcal{U}_{\rm MF}+
\Sigma_{\rm MF}(\overline{q}q-\langle\overline{q}q\rangle)\nonumber\\
\label{IX}
&+&G_S(\overline{q}q-\langle\overline{q}q\rangle)^2+G_{PS}(\overline{q}i\gamma_5\vec\tau q)^2.
\end{eqnarray}
Hereafter the subscript index ``$\rm MF$'' labels the quantities defined at the mean-field. Eq. (\ref{IX}) includes only three non-vanishing expansion coefficients 
\begin{eqnarray}
\label{X}
\Sigma_{\rm MF}&=&\frac{\partial\mathcal{U}_{\rm MF}}{\partial\langle\overline{q}q\rangle},\\
\label{XI}
G_S&=&-\frac{1}{2}
\frac{\partial^2\mathcal{U}_{\rm MF}}{\partial\langle\overline{q}q\rangle^2},\\
\label{XII}
G_{PS}&=&-\frac{1}{6}
\frac{\partial^2\mathcal{U}_{\rm MF}}{\partial\langle\overline{q}i\gamma_5\vec\tau q\rangle^2}.
\end{eqnarray}
Substituting $\mathcal{U}$ in Eq. (\ref{I}) with $\mathcal{U}^{(2)}$ the model Lagrangian can be given an effective current-current interaction form of the NJL model type. 
In this case $\Sigma_{\rm MF}$ interpreted as the mean-field self-energy of quarks is absorbed to their effective mass $m^*=m+\Sigma_{\rm MF}$, while $G_{S}$ and $G_{PS}$ turn out to be the effective couplings in the scalar and pseudoscalar interaction channels, respectively. 
In the vacuum the mean-field self-energy of quarks $\Sigma_{\rm MF}\propto\alpha^{-2/3}$ diverges at vanishing $\alpha$ and attains a finite value at $\alpha\neq0$. In other words, this parameter controls a phenomenological suppression of quark degrees of freedom in the confining region by high or even divergent values of their mass. The role of $D_0$ becomes clear from the consideration of the mean-field self-energy of heavy quarks within the so-called no-sea approximation when the chiral condensate is the opposite of the quark number density, i.e. $\langle\overline{q}q\rangle=-\langle q^+q\rangle$ \cite{Kaltenborn:2017hus}. In this case $\Sigma_{\rm MF}$ is proportional to the mean separation between quarks $\langle q^+q\rangle^{-1/3}$, which allows us to interpret it as energy associated with an extended string joining two quarks and characterized by the string tension $\propto D_0$.

In the general case, medium-dependent $G_S$ and $G_{PS}$ differ signaling about violation of chiral symmetry. 
This happens as a result of expanding the potential $\mathcal{U}$ around the chirally broken mean-field solution. 
On the other hand, dynamical restoration of chiral symmetry at high densities drives $G_{S}$ and $G_{PS}$ to the same asymptotic value \cite{Ivanytskyi:2021dgq,Ivanytskyi:2022oxv,Ivanytskyi:2022bjc}, which is close to the one of the NJL model \cite{Ratti:2005jh}.

In Ref. \cite{Ivanytskyi:2022oxv} scalar and pseudoscalar correlations of quarks caused by interaction in the corresponding channels were analyzed within the Gaussian approximation in order to extract mass $M_\pi$ and decay constant $F_\pi$ of pion, as well as the mass of the scalar meson $M_\sigma$. 
The state $f_0(980)$ was considered the most credible candidate for the role of scalar meson due to a rather high width of about 500-1000 MeV of the lighter state $f_0(500)$, which experimental status remains unclear \cite{PhysRevD.98.030001}. 
The vacuum value of the chiral condensate per flavor $\langle\overline{l}l\rangle_0$ is another quantity important for the QCD phenomenology. 
Similarly to most chiral models of quark matter \cite{Grigorian:2006qe} our approach is unable to reproduce $|\langle\overline{l}l\rangle^{1~GeV}_0|^{1/3}=241$ MeV obtained from QCD sum rules at the renormalization scale 1 GeV \cite{Jamin:2002ev}. 
Due to this, analysis of $\langle\overline{l}l\rangle_0$ within the present model was performed together with the pseudocritical temperature $T_{PC}$ defined by the peak position of chiral susceptibility at vanishing baryon density. Thus, $M_\pi=140$ MeV, $F_\pi=90$ MeV, $M_\sigma=980$ MeV, $|\langle\overline{l}l\rangle_0|^{1/3}=267$ MeV and $T_{PC}=163$ MeV were used to fit four model parameters, i.e. the current quark mass $m$, the interaction potential parameters $D_0$ and $\alpha$, as well as momentum scale $\Lambda$, which regularizes zero point terms in the expression for the thermodynamic potential (see Refs. \cite{Ivanytskyi:2022oxv,Ivanytskyi:2022bjc} for details). The values given in Table \ref{table1} yield $m^*_0=718$ MeV in the vacuum, which provides efficient phenomenological confinement of quarks at low temperatures and densities.

\begin{table}[t]
\begin{tabular}{|c|c|c|c|c|c|c|c|c|c|c|c|}
\hline
$m$ [MeV] & $\Lambda$ [MeV] & $\alpha$ & $D_0\Lambda^{-2}$ & $M_g$ [MeV]  \\ \hline 
    4.2   &       573       &   1.43   &      1.39         &      600     \\ \hline 
\end{tabular}
\caption{Values of the RDF approach parameters.}
\label{table1}
\end{table}

Within the present approach vacuum values of the vector $G_{V0}$ and diquark $G_{D0}$ couplings are treated as free parameters parameterized by the dimensionless ratios 
\begin{eqnarray}
\label{eq:couplings}
\eta_V\equiv \frac{G_{V0}}{G_{S0}}
\quad{\rm and}\quad 
\eta_D\equiv \frac{G_{D0}}{G_{S0}}
\end{eqnarray}
with  $G_{S0}=18.1~{\rm GeV}^{-2}$ being the scalar coupling in the vacuum.
The dimesionless diquark coupling $\eta_D$ controls the position of the propagator pole of the auxiliary diquark field.
This field appears within the RDF approach as a result of the bosonization of $\mathcal{L}_D$ via the Hubbard-Stratonovich transformation (see Ref. \cite{Ivanytskyi:2022oxv} for details).
The diquark propagator pole gives access to the diquark mass.
At too large values of $\eta_D>\eta_D^{\rm max}$, the diquark mass vanishes in the vacuum and thus coincides with the corresponding chemical potential \cite{Sun:2007fc,Zablocki:2009ds}.
In this case, Bose-Einstein condensation of massless diquarks and antidiquarks occurs, corresponding to the formation of a color-superconducting vacuum state with a finite value of the diquark condensate.
This would entail that the global minimum of the vacuum thermodynamic potential $\Omega_0$ would occur at a nonvanishing value of $\langle\overline{q}^ci\tau_2\gamma_5\lambda_2 q\rangle$, while the normal vacuum state with $\langle\overline{q}^ci\tau_2\gamma_5\lambda_2 q\rangle=0$ would correspond to a local maximum so that the normal vacuum state would become unstable against a transition to the color-superconducting one.
Since the physical QCD vacuum is not color superconducting, one has to require that physical values of the diquark coupling be limited by the condition $\eta_D<\eta_D^{\rm max}=(3/2)(G_{PS0}/G_{S0})m^*_0/(m^*_0-m)$ \cite{Ivanytskyi:2022oxv,Ivanytskyi:2022bjc}.
The model parameters from Table \ref{table1} yield $\eta_D^{\rm max}=0.78$.
This value does not depend on $\eta_V$ and $M_g$ since quark number density vanishes in the vacuum.

In this work, we analyze the intervals $\eta_V\in[0.1,0.4]$ and $\eta_D\in[0.2,0.8]$, which include the values of the vector and diquark couplings providing the best agreement with the observational constraints on the mass-radius relation and tidal deformability of NSs \cite{Ivanytskyi:2022oxv}.
Note, unphysical values of $\eta_D>\eta_D^{\rm max}$ are included only for the sake of having the widest interval of the corresponding parameter needed to increase the quality of mapping $\eta_V$ and $\eta_D$ to the ABPR parameters.
Astrophysical applications presented in this work are limited to $\eta_D<\eta_D^{\rm max}$.

\begin{figure}[ht]
\includegraphics[width=\columnwidth]{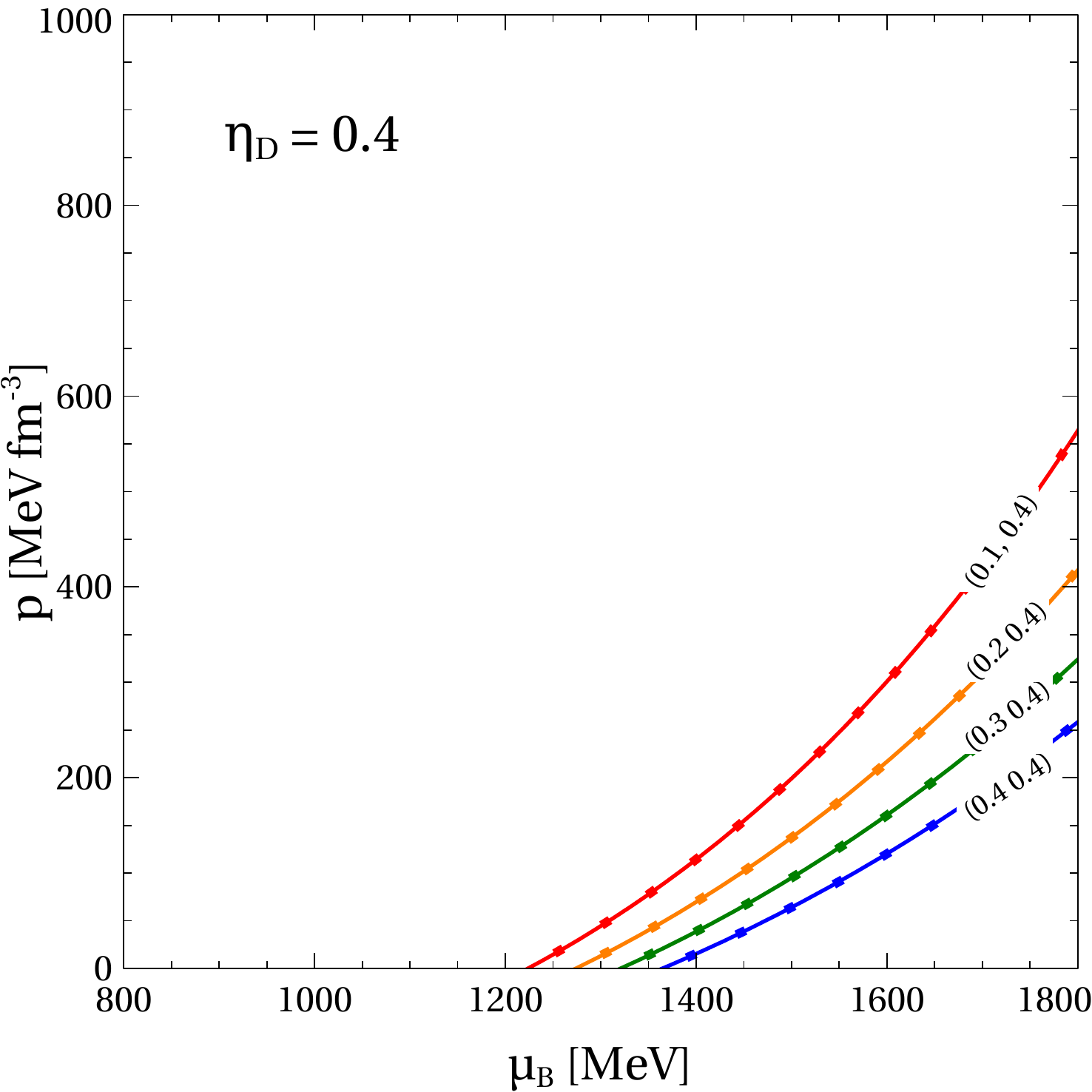}
\includegraphics[width=\columnwidth]{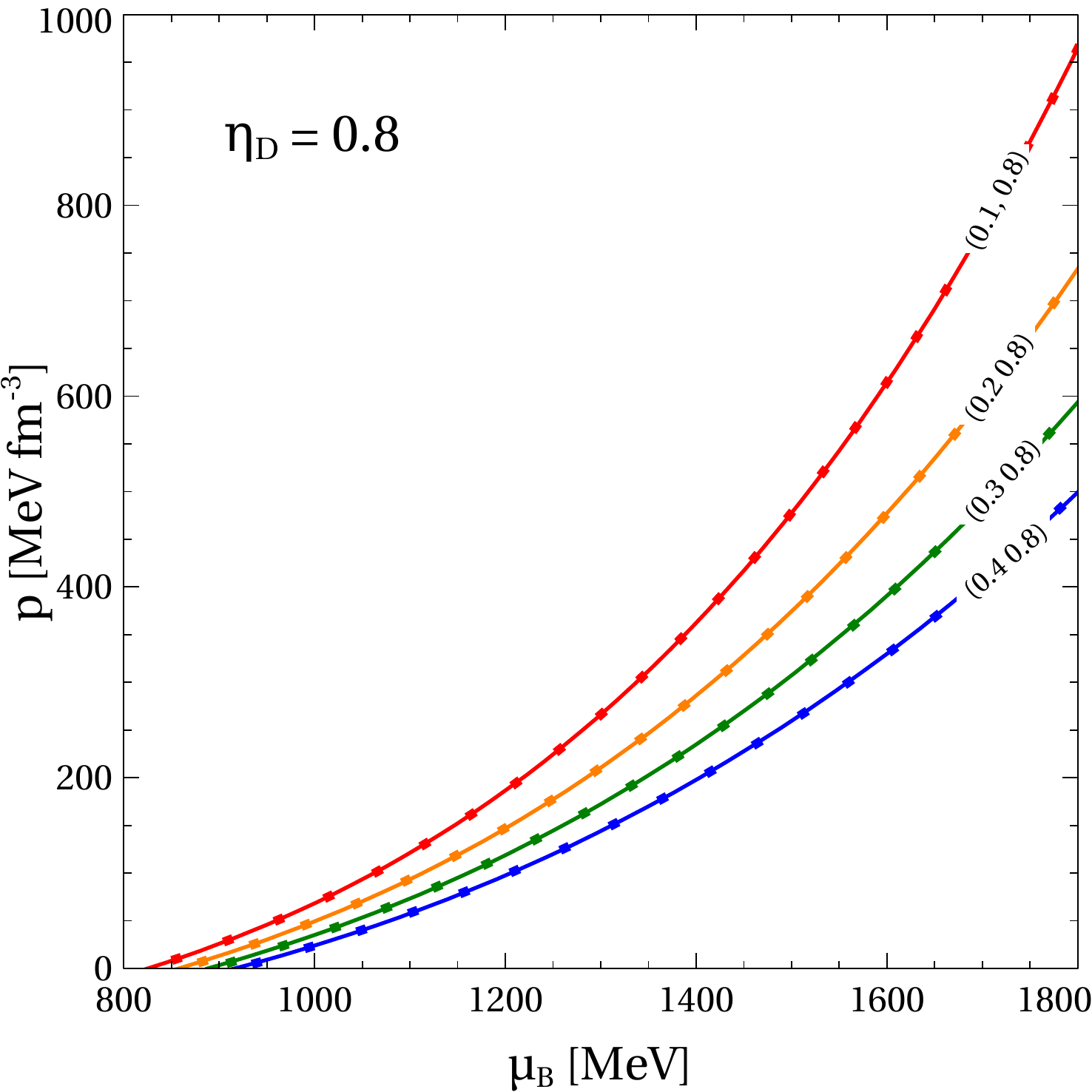}
\caption
{Pressure $p$ of cold electrically neutral quark matter at $\beta$-equilibrium as a function of baryonic chemical potential $\mu_B$, calculated for $\eta_D=0.4$ (upper panel), $\eta_D=0.8$ (lower panel) and several values of $\eta_V$. 
The solid curves labeled with pairs of numbers $(\eta_V,\eta_D)$ correspond to the RDF approach, while the square shaped dots depict the ABPR model EoS (\ref{XIII}) with parameters from Table \ref{table2}, respectively. 
}
\label{fig1}
\end{figure}

For given $\eta_V$ and $\eta_D$ the EoS of quark matter is obtained by applying the mean-field approximation to the Lagrangian (\ref{I}) with $\mathcal{U}$ substituted by $\mathcal{U}^{(2)}$ (see Ref. \cite{Ivanytskyi:2022bjc} for details).
In this work, we consider the case of cold quark matter at $\beta$-equilibrium. Electric neutrality of the NS matter is guaranteed by the proper amount of electrons. The pairs of numbers $(\eta_V,\eta_D)$ are used to label different EoSs. 
Instead of a detailed discussion, which can be found in Refs. \cite{Ivanytskyi:2022oxv,Ivanytskyi:2022bjc}, here we just briefly outline these EoSs. 
In the vacuum and at small values of the baryonic chemical potential $\mu_B$ chiral symmetry is broken, the chiral condensate has its large vacuum value $\langle\overline{q}q\rangle=\langle\overline{q}q\rangle_0$, and quarks have a large mass coinciding with its vacuum value. This suppresses quark excitation and leads to vanishing pressure $p$, quark number density $\langle q^+ q\rangle$, diquark condensate $\langle\overline{q}^ci\gamma_5\tau_2\lambda_2 q\rangle$ and energy density $\varepsilon$. 
Quark pairing gap $\Delta=2G_D|\langle\overline{q}^ci\gamma_5\tau_2\lambda_2 q\rangle|$ and squared speed of sound $c_S^2=dp/d\varepsilon$ also vanish at this regime.
At a certain value of $\mu_B$ chiral condensate and quark mass drop to some small values $\langle\overline{q}q\rangle/\langle\overline{q}q\rangle_0<1$ and $m^*/m^*_0<1$. This manifests the first order phase transition, at which $\langle q^+ q\rangle$, $\langle\overline{q}^ci\gamma_5\tau_2\lambda_2 q\rangle$ and $\varepsilon$ discontinuously attain some finite values, while $p$ remains continuous. Above this transition, quark matter exists in the color superconducting 2SC phase, since $\Delta\neq0$ in this case. 
Fig. \ref{fig1} shows the pressure of the present RDF approach as a function of baryonic chemical potential for two limiting values of the considered interval of the diquark coupling and several values of the vector coupling belonging to the interval considered in this paper.

\subsection{The ABPR EoS}
\label{sec2_2}

\begin{table}[t]
\begin{tabular}{|c|c|c|c|c|c|}
\hline
\multicolumn{6}{|c|}{$A_4$} \\ \hline
\diaghead{\hspace*{1.5cm}}{$\eta_V$}{$\eta_D$} 
     &  0.4  &  0.5  &  0.6  &  0.7  &  0.8  \\ \hline
 0.1 & 0.738 & 0.781 & 0.819 & 0.857 & 0.898 \\ \hline
 0.2 & 0.501 & 0.521 & 0.540 & 0.555 & 0.571 \\ \hline
 0.3 & 0.369 & 0.379 & 0.389 & 0.394 & 0.403 \\ \hline
 0.4 & 0.281 & 0.289 & 0.296 & 0.299 & 0.303 \\ \hline 
                                                \hline
 \multicolumn{6}{|c|}{$\Delta~[{\rm MeV}]$}  \\ \hline
 \diaghead{\hspace*{1.5cm}}{$\eta_V$}{$\eta_D$}  
     &  0.4  &  0.5  &  0.6  &  0.7  &  0.8   \\ \hline
 0.1 & 162.6 & 185.4 & 209.5 & 231.0 & 248.0  \\ \hline
 0.2 & 201.9 & 220.7 & 238.9 & 256.9 & 273.0  \\ \hline
 0.3 & 214.0 & 230.4 & 245.7 & 262.2 & 275.1  \\ \hline
 0.4 & 219.1 & 233.0 & 246.1 & 260.1 & 271.7  \\ \hline
                                                 \hline 
\multicolumn{6}{|c|}{$B~[{\rm MeV/fm}^3]$}    \\ \hline
\diaghead{\hspace*{1.5cm}}{$\eta_V$}{$\eta_D$} 
     &  0.4  &  0.5  &  0.6  &  0.7  &  0.8   \\ \hline
 0.1 & 192.5 & 185.3 & 168.2 & 138.8 & 95.0   \\ \hline
 0.2 & 204.0 & 196.9 & 178.6 & 148.9 & 105.4  \\ \hline
 0.3 & 208.1 & 200.8 & 181.9 & 152.4 & 108.6  \\ \hline
 0.4 & 210.1 & 202.5 & 183.1 & 153.1 & 108.8  \\ \hline
\end{tabular}
\caption{Values of the parameters $A_4$, $\Delta$ and $B$ of the two-flavor ABPR EoS fitted to the microscopic RDF approach with the corresponding values of $\eta_V$ and $\eta_D$ as described in the text.}
\label{table2}
\end{table}

\begin{figure}[t]
\includegraphics[width=\columnwidth]{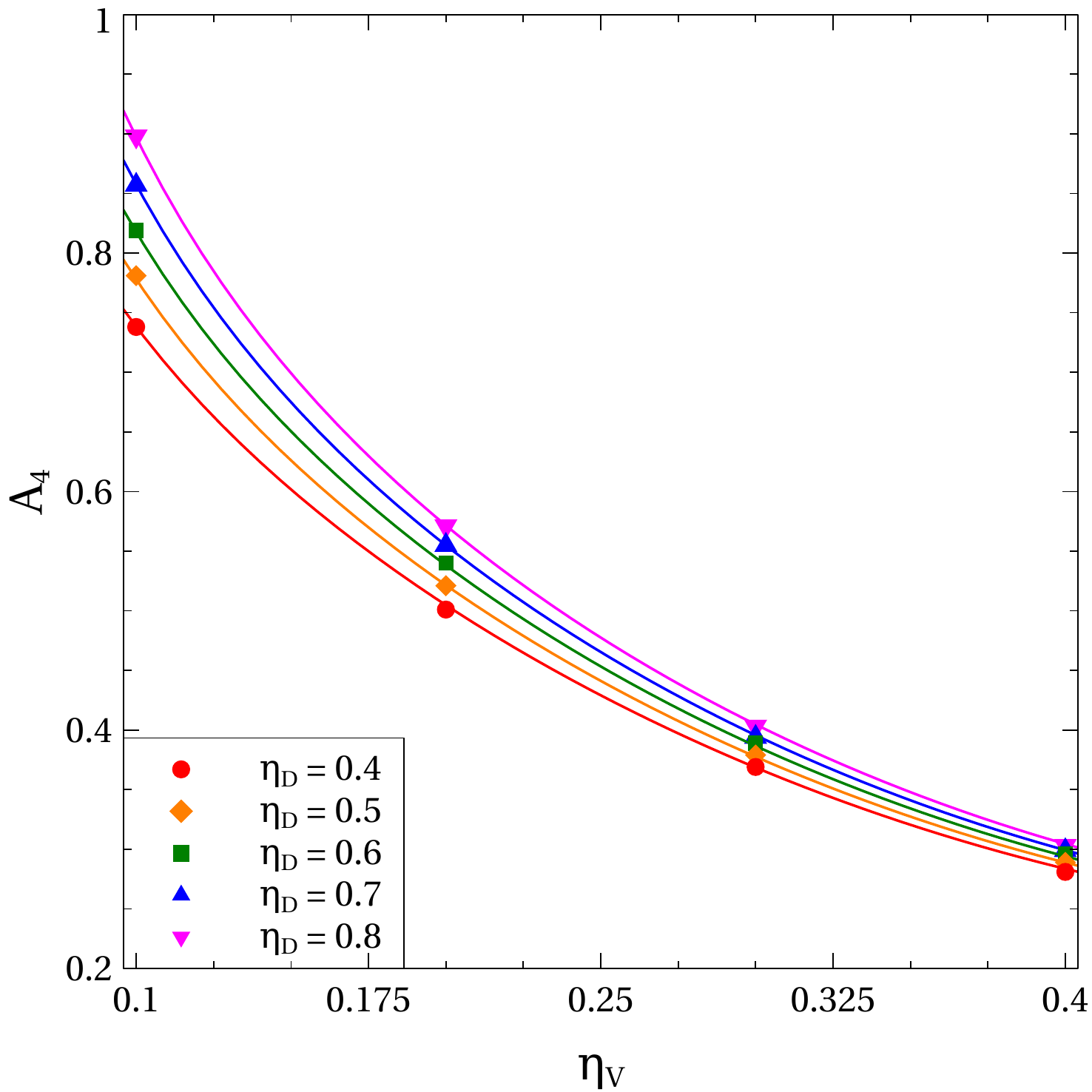}
\includegraphics[width=\columnwidth]{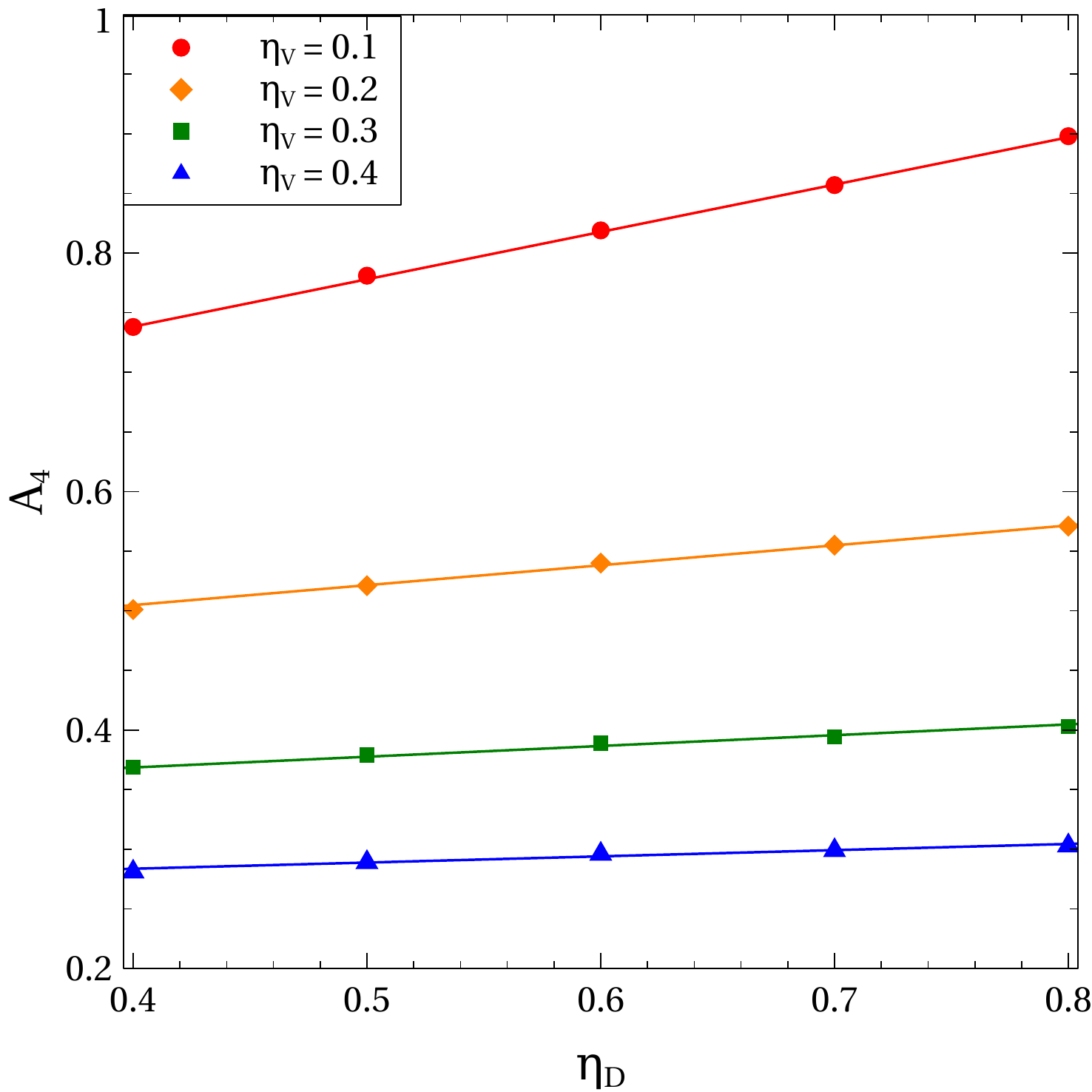}
\caption
{The ABPR EoS parameter $A_4$ as a function of $\eta_D$ (upper panel) and $\eta_V$ (lower panel).
Filled circles, diamonds, squares, and triangles represent the values from Table \ref{table2}, while solid lines correspond to the functional dependence given by Eq. (\ref{XIV}).
}
\label{fig2}
\end{figure}

\begin{figure}[t]
\includegraphics[width=\columnwidth]{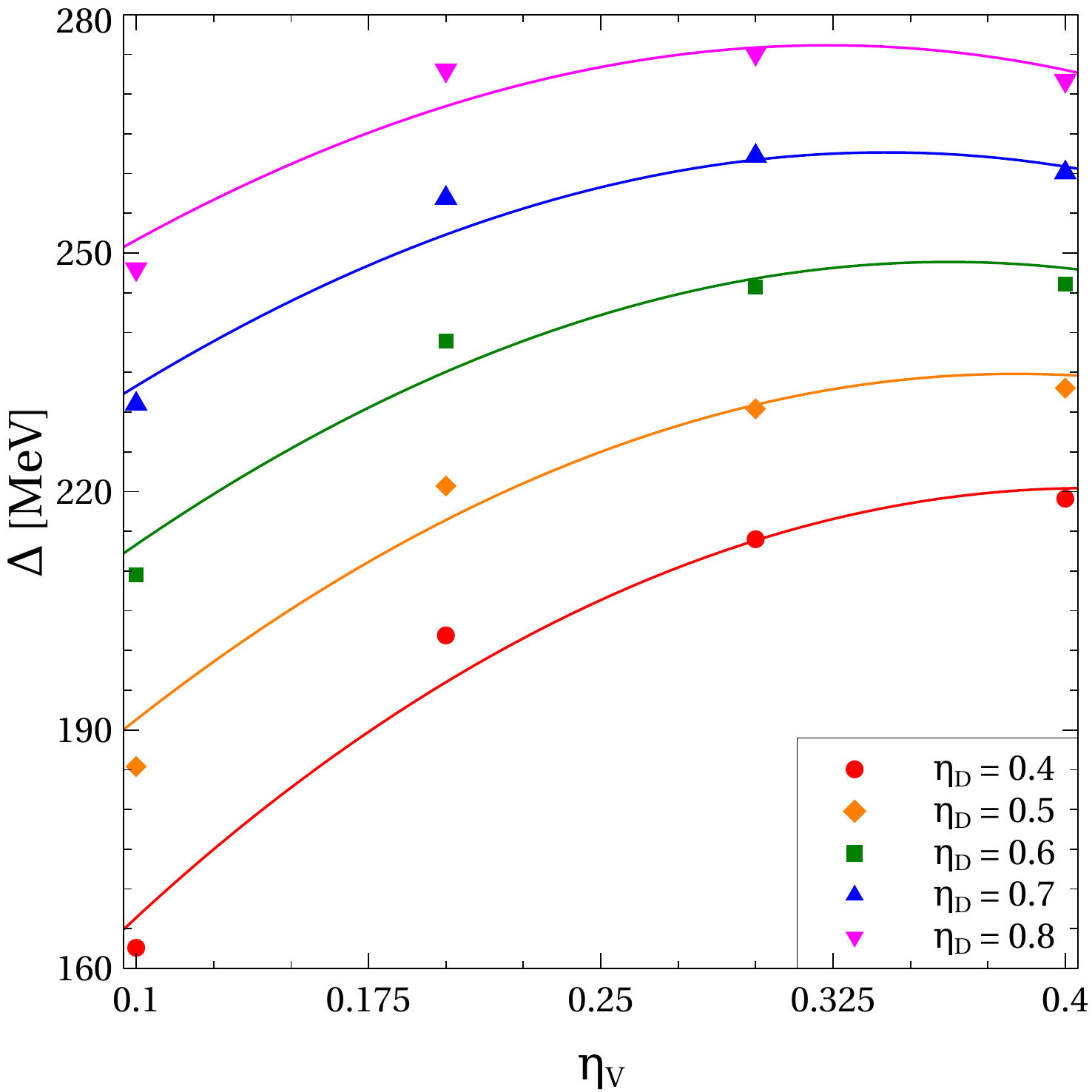}
\includegraphics[width=\columnwidth]{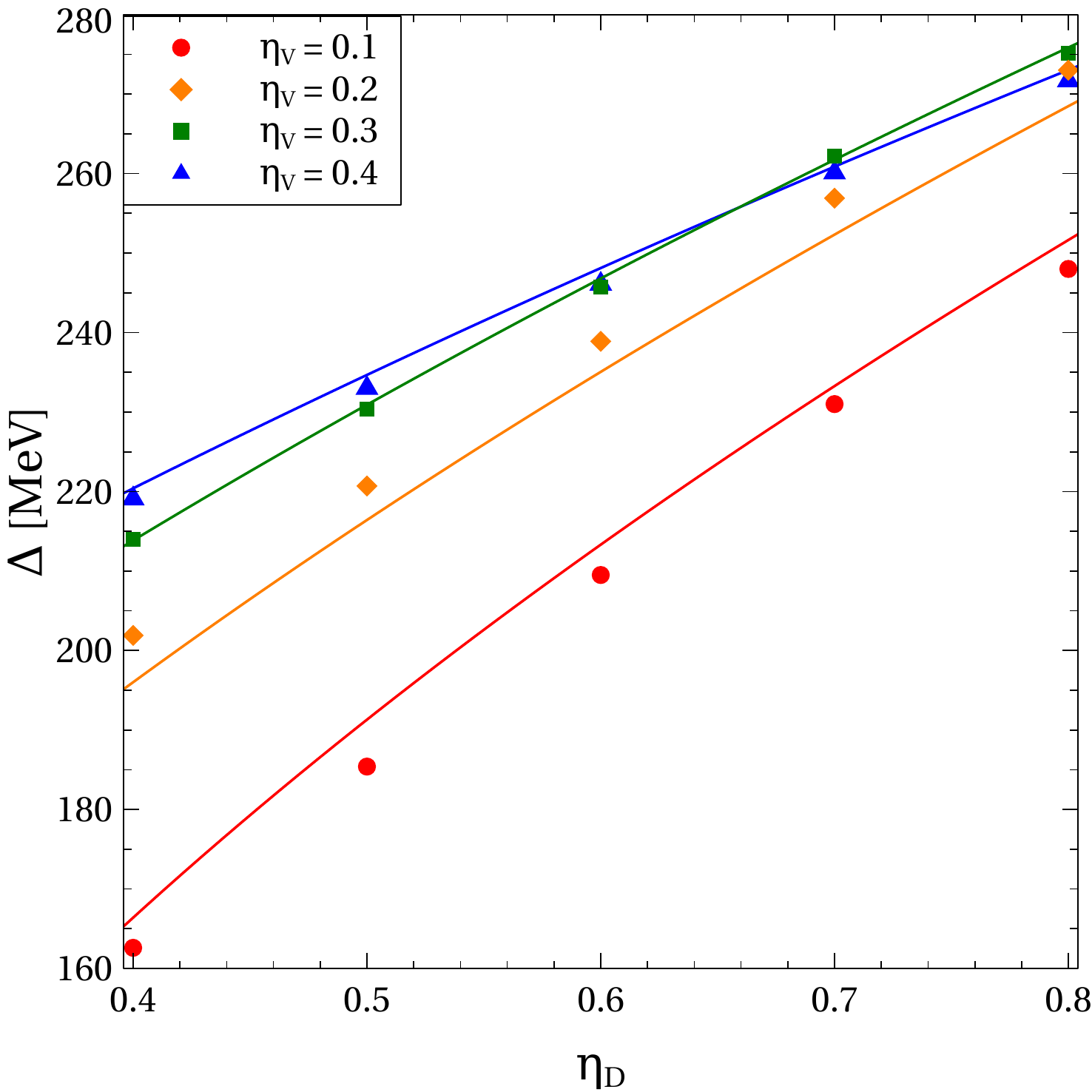}
\caption
{The same as in Fig. \ref{fig2} but for the diquark gap $\Delta$ parameterized by Eq. (\ref{XV}).
}
\label{fig3}
\end{figure}

\begin{figure}[t]
\includegraphics[width=\columnwidth]{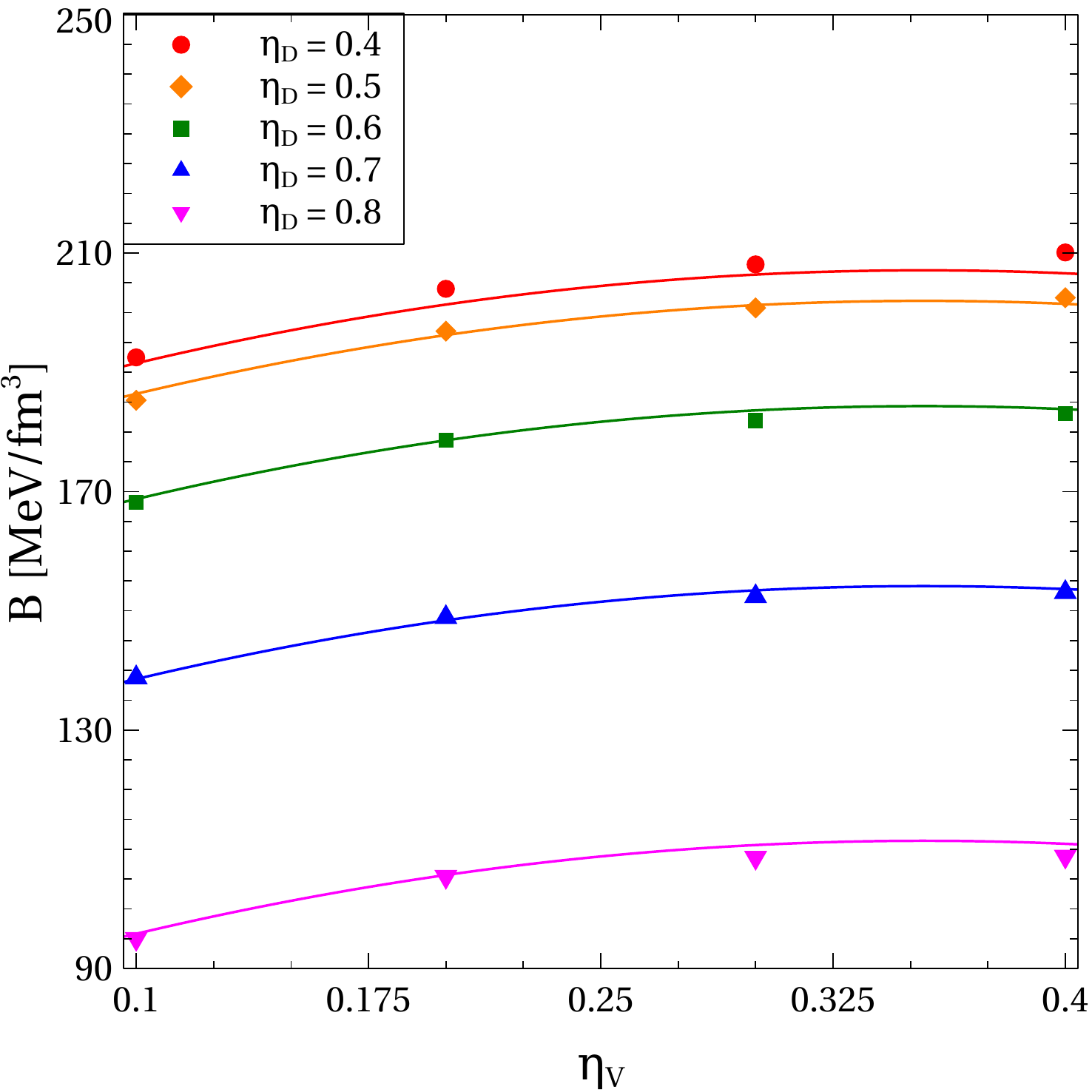}
\includegraphics[width=\columnwidth]{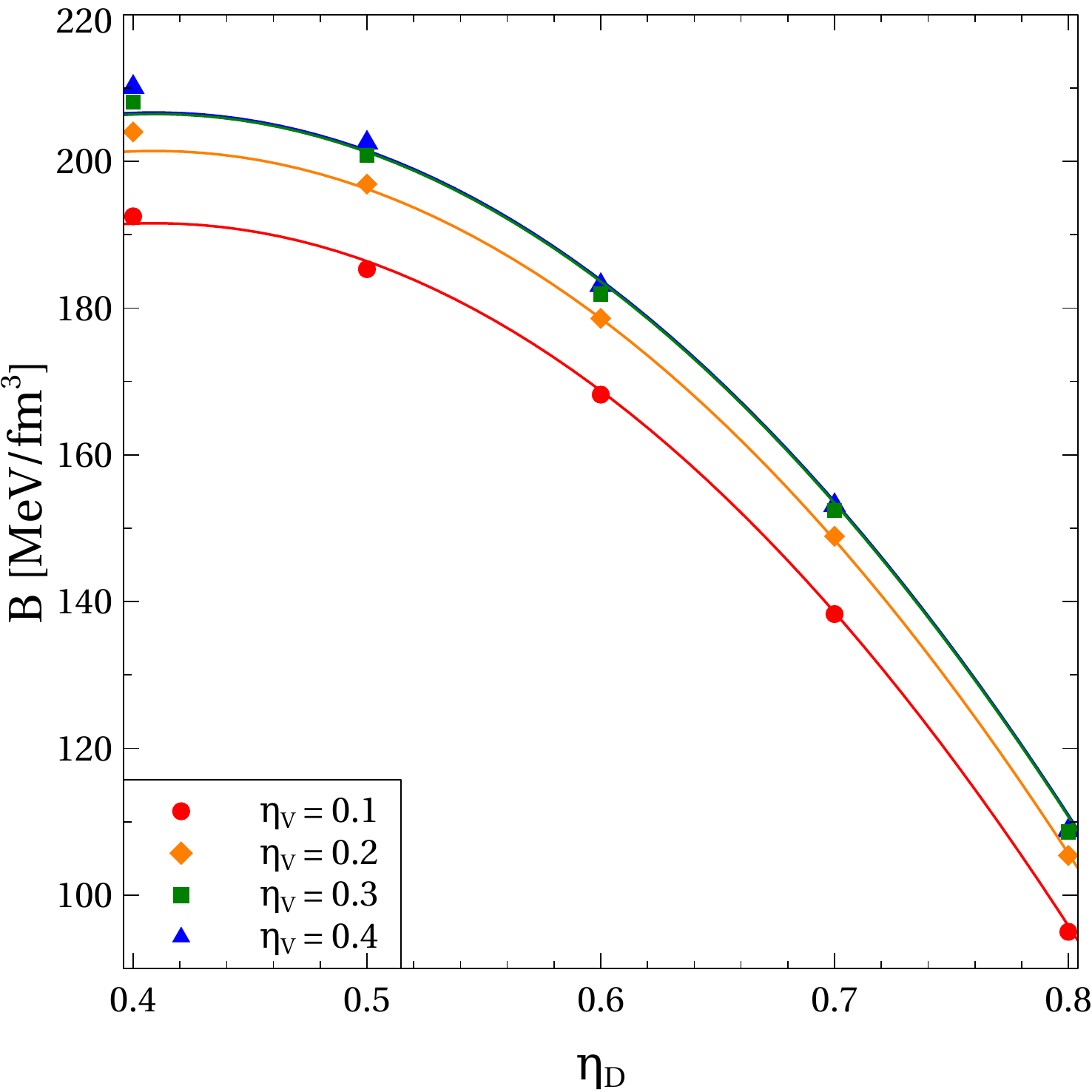}
\caption
{The same as in Fig. \ref{fig2} but for the bag pressure $B$ parameterized described by Eq. (\ref{XVI}).
}
\label{fig4}
\end{figure}

The ABPR model represents a phenomenological extension of the bag model of three-flavor quark matter, which accounts for the perturbative QCD correction to pressure and effects of quark pairing \cite{Alford:2004pf}.
Recently it got a microscopic justification based on a version of the non-local NJL model with three degenerate quark flavors \cite{Blaschke:2022egm}.
In this case, electric neutrality is provided even in the absence of electrons.
The pressure of the original three-flavor ABPR model as a function of quark chemical potential $\mu=\mu_B/3$ reads 
\begin{eqnarray}
\label{XIIIa}
p=\frac{3}{4\pi^2}A_4\mu^4
+\frac{3}{\pi^2}\left(\Delta^2-\frac{m_s^2}{4}\right)\mu^2-B.
\end{eqnarray}
The first term in this expression is proportional to the number of $2\times3\times3$ spin-color-flavor quark states.
The parameter $A_4=1-2\alpha_s/\pi$ absorbs the order $\mathcal{O}(\alpha_s)$ perturbative correction mentioned above. 
Hereafter $\alpha_s=g^2/4\pi$ is the QCD fine structure constant expressed through the QCD running coupling $g$ \cite{Kapusta:2021ney}.
In the two-flavor case, this term should be modified by the factor $2/3$, while $A_4$ should absorb not only the perturbative correction but also the pressure of electrons. 
Therefore, simple relation $A_4=1-2\alpha_s/\pi$ does not hold in the two-flavor case.
Due to this, we consider $A_4$ as a free parameter.
The second term of the three-flavor ABPR EoS (\ref{XIIIa})
contains the effective constant diquark pairing gap $\Delta$, which represents the correction corresponding to quark pairing, and the strange quark mass $m_s$.
In this work, we consider the two-flavor case, where strange quarks are absent and therefore no strange quark mass term appears in the two-flavor ABPR EoS given below.
At three quark flavors, the quark pairing correction is also proportional to the number of $2\times3\times3$ spin-color-flavor quark states, since all of them are paired in this case.
At the same time, these states split into singlet and octet ones characterized by the pairing gaps $\Delta$ and $2\Delta$, respectively.
This produces an additional factor in the quark pairing term of the ABPR EoS applied to the three flavor quark matter, i.e. $1^2\times8/9+2^2\times1/9=4/3$.
However, in the two-flavor case, only two of the colored states pair, while their splitting to the singlet and octet states is absent. 
Thus, the second term of the three-flavor ABPR EoS should be modified by the factor $2\times2/(3\times3\times4/3)=1/3$.
The third term $-B$ stands for the bag pressure assuring the confining property of the quark EoS.
Thus, in the two-flavor case, we arrive at 
\begin{eqnarray}
\label{XIII}
p=\frac{A_4\mu^4}{2\pi^2}+\frac{\Delta^2\mu^2}{\pi^2}-B.
\end{eqnarray}
Baryon number density, energy density, and squared speed of sound of this EoS can be found using the thermodynamic identities $n_B=\partial p/\partial\mu_B$, $\varepsilon=\mu_B n_B-p$ and $c_S^2=dp/d\varepsilon$, respectively.
Its high-density behavior is conformal, since $p\rightarrow A_4\mu^4/2\pi^2$ and $c_S^2\rightarrow1/3$ at $\mu\rightarrow\infty$.

The parameters $A_4$, $\Delta$, and $B$ can be fixed by fitting this simple phenomenological EoS to the microscopic RDF approach.
In this study, such a fit is performed for the interval of baryonic chemical potentials limited by $\mu_B<1800$ MeV, which covers the range relevant for the NS interior \cite{Ivanytskyi:2022bjc} but does not go far beyond it.
Our analysis evidences that at these values of $\mu_B$, Eq. (\ref{XIII}) perfectly fits the EoS of the RDF approach obtained for wide ranges of the vector and diquark couplings.
Fig. \ref{fig1} demonstrates a perfect agreement between the ABPR EoS and the RDF approach at the limiting values of $\eta_D$ and $\eta_V=0.1 - 0.4$.
The same agreement is reached for any $\eta_D$ and $\eta_V$ from the intervals mentioned above. 
Values of the ABPR EoS parameters defined for different vector and diquark couplings are given in Table \ref{table2}.
Figs. \ref{fig2}, \ref{fig3} and \ref{fig4} show the corresponding dependences.

\subsection{Parameter matching between the ABPR EoS and RDF approach}
\label{sec2_3}

Here we show how the quark matter EoS can be fitted by the ABPR model that gives a simple functional dependence between the parameters of this phenomenological EoS and microscopic parameters of the initial Lagrangian, i.e. couplings $\eta_V$ and $\eta_D$. 
This result could be further utilized to perform numerical-relativity simulations within this model or other studies.

The dependence of $\Delta$ and $B$ on $\eta_V$ can be reproduced by a second-order polynomial. 
At the same time, $A_4$ exhibits rather fast growth at small values of the vector coupling, which requires a correction $\propto1/\eta_V$. 
Furthermore, it is seen from Fig.~\ref{fig2} that $A_4$ is perfectly linear in the diquark coupling.
On the other hand, $\Delta$ and $B$ are concave functions of $\eta_D$.
While the mild $\eta_D$-dependence of the parameter $\Delta$ extracted from fitting the RDF EoS can be described by a square root behavior, the bag pressure is more curved and requires a parabolic scaling.
These aspects of the ABPR parameters behavior are accounted for by the parameterization 
\begin{eqnarray}
\label{XIV}    
A_4&=&a_1+b_1\eta_{V}+c_1\eta_{V}^{2}+\left(d_1+\frac{e_1}{\eta_{V}}\right)\eta_D,\\
\label{XV}
\Delta&=&(a_2+b_2\eta_V+c_2\eta_V^2)\sqrt{d_2+e_2\eta_V+\eta_D},\\
\label{XVI}  
B&=&a_3+b_3\eta_V+c_3\eta_{V}^{2}+d_3\eta_{D}+e_3\eta_{D}^{2}.
\end{eqnarray}
They are given in terms of the fifteen constant parameters $a_i$, $b_i$, $c_i$, $d_i$, and $e_i$ with $i=1,2,3$.
Their values can be extracted by fitting Eqs. (\ref{XIV}) - (\ref{XVI}) to the values of $A_4$, $\Delta$ and $B$ from Table \ref{table2}. 
The results of such a fit are listed in Table \ref{table3}.

\begin{table}[t]
\begin{tabular}{|c|c|c|c|c|c|c|c|c|c|c|c|}
\hline
i &       units      & $a_i$ & $b_i$ & $c_i$ & $d_i$ & $e_i$  \\ \hline
1 &                  & 0.757 &-1.955 & 1.799 &-0.063 & 0.046  \\ \hline
2 &       [MeV]      & 300.7 & 8.534 &-308.2 &-0.235 & 1.458  \\ \hline
3 & $\rm[MeV/fm^3]$  & 72.018& 170.8 &-241.0 & 512.7 &-626.6  \\ \hline
4 &  $\rm[M_\odot]$  & 30.470&-130.8 & 463.3 &  $-$  &  $-$   \\ \hline
5 &  $\rm[M_\odot]$  &-40.213& 165.5 &-581.9 &  $-$  &  $-$   \\ \hline
\end{tabular}
\caption{Values of the parameters of Eqs. (\ref{XIV}) - (\ref{XVI}), (\ref{eq:Monset_fit}). The symbol ``$-$'' indicates that the corresponding parameter does not exist.}
\label{table3}
\end{table}

Eqs. (\ref{XV})-(\ref{XVI}) present the RDF EoS in the ABPR form.
This representation allows us to discuss the relation of this microscopic EoS to the asymptotics of the order $\mathcal{O}(\alpha_s)$ perturbative QCD in the spirit of Ref. \cite{Blaschke:2022egm}.
For this, we notice that the chemical potential of the electrons $\mu_e$ scales under the conditions of electric neutrality and $\beta$-equilibrium with a two-flavor gas of free massless quarks as $\mu_e=0.219\, \mu$,
while the chemical potentials of quarks are $\mu_u=0.854\, \mu$ and $\mu_d=1.073\, \mu$.
This yields a free quark-electron pressure 
$p_{\rm free}=0.929\,\mu^4/2\pi^2$.
The $\mathcal{O}(\alpha_s)$ perturbative correction to this pressure is $p_{\rm pert}=-\alpha_s(\mu_u^4+\mu_d^4)/2\pi^3=-1.857\,\alpha_s\mu^4/2\pi^3$.
This allows us to relate the QCD fine structure constant to the $A_4$ parameter of the two-flavor ABPR EoS given by Eq. (\ref{XIII}) as $A_4=2\pi^2(p_{\rm free}+p_{\rm pert})/\mu^4=0.929-1.854\,\alpha_s/\pi$.
In Section \ref{sec3} we argue that small values of $\eta_V<0.2$ are inconsistent with the present observational constraints on the mass-radius relation of NSs.
As is seen from the upper panel of Fig. \ref{fig2}, this suggests $A_4<0.6$ or, equivalently, $\alpha_s>0.557$.
Due to such large values of the extracted fine structure constant of QCD we conclude that quark matter in the NS cores is far from the perturbative regime.

\section{Hybrid stars with quark core}
\label{sec3}

\begin{figure}[t]
\includegraphics[width=\columnwidth]{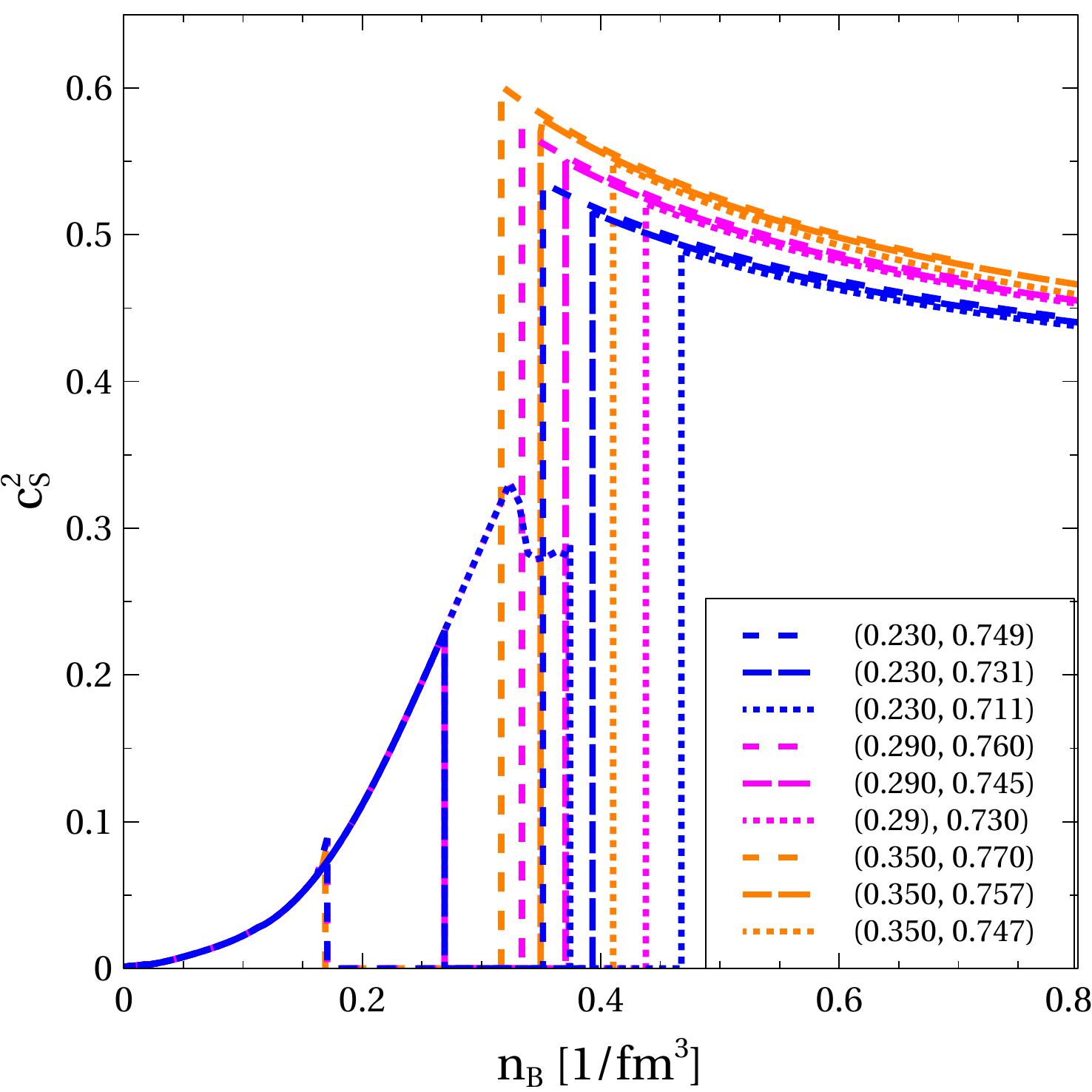} 
\caption{
Squared speed of sound as a function of the baryon density for the hybrid EoS with a first-order phase transition resulting from a Maxwell construction with the hadronic DD2npY-T EoS \cite{Shahrbaf:2022upc} and the ABPR form of the RDF EoS of quark matter (\ref{XIII})-(\ref{XVI}) calculated for three values of the vector coupling $\eta_V=0.230$ (blue lines), 0.290 (magenta lines) and 0.350 (orange lines) and the diquark coupling, which lead to the quark onset densities $n_{\rm onset} =0.17~{\rm fm}^{-1}$ (short dashed lines), $0.27~{\rm fm}^{-1}$ (long dashed lines) and $0.37~{\rm fm}^{-1}$ (dotted lines).}
\label{fig:cs2nB}
\end{figure}

\subsection{Hybrid EoS and TOV solutions}
Eqs. (\ref{XIV}) - (\ref{XVI}) inserted to Eq. (\ref{XIII}) represent a thermodynamically equivalent ABPR representation of the RDF approach to quark matter.
This representation is given in terms of the microscopic parameters of the RDF approach, i.e. coupling constants.
In this section, we utilize it to model NSs with quark cores.
Outer layers of such stars consist of hadronic matter modeled by the DD2npY-T EoS \cite{Shahrbaf:2022upc}. 
This RDF EoS includes nucleonic and hyperonic degrees of freedom and agrees with the low-density constraint from the chiral effective field theory \cite{Kruger:2013kua}.
The quark and hadron parts of the hybrid quark-hadron EoS are matched via the Maxwell construction.
In Fig. \ref{fig:cs2nB}, we show the squared speed of sound as a function of the baryon density for the resulting hybrid EoSs.
As is seen, the speed of sound vanishes in the mixed quark-hadron phase since pressure is constant there.
This absence of a pressure gradient is exactly the reason why a Maxwell-constructed mixed phase is not realized as a solution of the TOV equations  (\ref{eq:Pprime}) and (\ref{eq:mprime}) for spherically symmetric, nonrotating NSs.
Indeed, using the thermodynamic identity $\mu_B=d\varepsilon/dn_B$ and the definition of the speed of sound $c_S^2=dp/d\varepsilon$, we can show that for the bulk modulus $K= n_Bdp/dn_B$ follows $K=\mu_Bn_Bc_s^2$.
Thus, the vanishing speed of sound leads to a vanishing bulk modulus and manifests the mechanical instability of the mixed phase against gravitational compression.
Furthermore, $c_S^2$ attains its maximal value of about $0.5-0.6$ right at the quark matter boundary of the mixed phase.
It exhibits a decrease with increasing density and reaches the conformal limit $c_S^2=1/3$ at asymptotically high $n_B$ \cite{Ivanytskyi:2022bjc}.

Constructing static spherically symmetric NSs with the hybrid EoS mentioned above requires solving the problem of relativistic hydrostatics in the Schwarzschild metric \citep{Schwarzschild:1916uq}
\begin{equation}
ds^2 = e^{\nu} dt^2 - e^{\lambda} dr^2 - r^2 (d \theta^2 + \sin^2 \theta d \phi^2),
\label{eq:metric}
\end{equation}
where $e^\nu$ and $e^{\lambda}$ are the metric functions.
In this metric, Einstein's field equations take the form of the TOV equations describing the interior solution of a nonrotating relativistic star in hydrostatic equilibrium \citep{Tolman:1939jz,Oppenheimer:1939ne},
\begin{eqnarray}
\dfrac{dp}{dr}& = & - (\varepsilon + p) \frac{{\rm m}+4 \pi p r^3}{r^2-2 r {\rm m}},
\label{eq:Pprime}
\\
\dfrac{d{\rm m}}{dr} & = & 4 \pi r^2 \varepsilon,
\label{eq:mprime}
\end{eqnarray}
where $\rm m$ is the gravitational mass enclosed in the sphere of radius $r$. 
The solution of these equations requires the knowledge of a relation connecting the pressure $p$ and energy density $\varepsilon$, i.e. the EoS of the hybrid NS matter.
The boundary conditions necessary to solve Eqs. (\ref{eq:Pprime}) and (\ref{eq:mprime}) are given by the central pressure $p_{\rm central}=p(r=0)$ (or, equivalently, the central baryonic chemical potential) and the fact that the gravitational mass enclosed in the very center of the star vanishes ${\rm m}_{\rm central}={\rm m}(r=0)=0$.

The solution of the TOV equations extends until the radial coordinate reaches the value $r={\rm R}$, at which the pressure $p$ vanishes, being equivalent to requiring dynamical equilibrium with the surrounding vacuum.
This $R$ corresponds to the NS radius, while the corresponding enclosed gravitational mass $\rm m=M$ is the NS mass. 
Beyond $r={\rm R}$ Eqs. (\ref{eq:Pprime}) and (\ref{eq:mprime}) are trivially satisfied by vanishing pressure and energy density and $\rm m=M$.
%
%

\begin{figure}[t]
\includegraphics[width=\columnwidth]{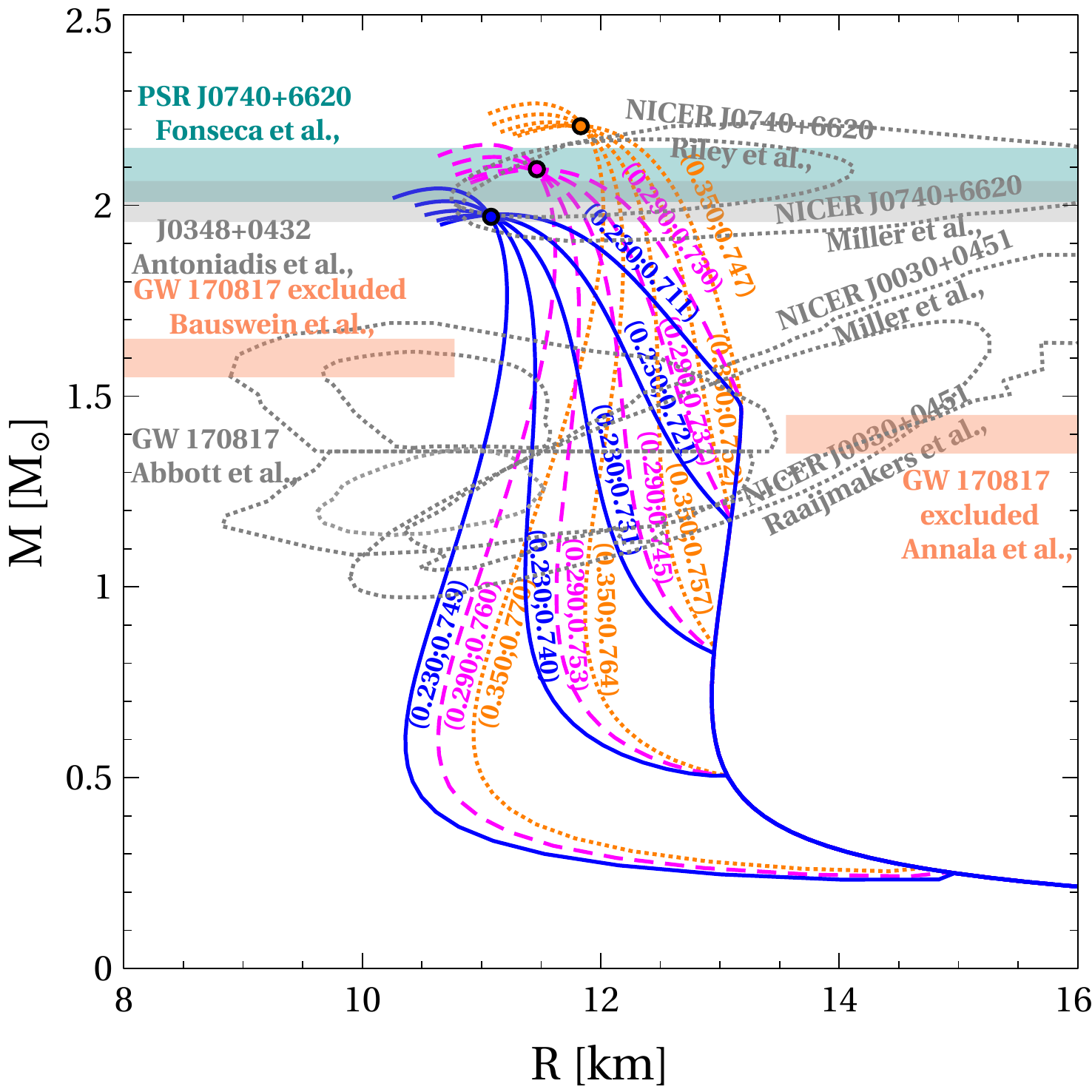} \caption{Mass-radius relation of hybrid stars with the
quark-hadron EoS constructed for the vector coupling $\eta_V=0.23$ (blue curves), $\eta_V=0.29$ (magenta curves), $\eta_V=0.35$ (orange curves), and several values of the diquark coupling. 
The curves are labeled with pairs of numbers ($\eta_V,\eta_D$).
Blue, magenta, and orange circles represent SPs, defined by the intersection of the curves of the same blue, magenta, and orange colors, respectively.
The astrophysical constraints depicted by the dotted contours and shaded bands are discussed in the text.
}\label{fig5}
\end{figure}

Fig. \ref{fig5} demonstrates the mass-radius relation of NSs with quark cores.
For modeling such objects we consider the values of the vector and diquark couplings, which provide consistency of the obtained hybrid quark-hadron EoSs with the observational constraints on $\rm M$ and $\rm R$.
These constraints include the lower limit of the TOV maximum mass $2.01{\pm 0.04}~\rm M_\odot$ measured in a binary white dwarf - pulsar PSR J0348+0432 system \cite{Antoniadis:2013pzd}, the limitations obtained from the Bayesian analysis of the observational data from the pulsars PSR J0740+6620 \cite{Riley:2021pdl,Miller:2021qha} and PSR J0030+0451 \cite{Riley:2019yda,Raaijmakers:2019qny,Miller:2019cac} and from the analysis of the gravitational wave signal from the NS merger GW170817 \cite{LIGOScientific:2018cki,Bauswein:2017vtn,Annala:2017llu}. 
We also confront the present model with the lower limit of the TOV maximum mass extracted from the continued timing observations of the PSR J0740+6620 radio-pulsar $2.08{\pm 0.07}~\rm M_\odot$\cite{Fonseca:2021wxt}.

It is seen from Fig. \ref{fig5} that with increasing the vector coupling the onset mass of quark deconfinement grows, while larger values of the diquark coupling lead to smaller values of the onset mass.
Fitting the values of $\rm M_{\rm onset}$ we find the empirical relation
\begin{eqnarray}
\label{eq:Monset_fit}    
{\rm M}_{\rm onset}=a_4+b_4\eta_V+c_4\eta_V^2+(a_5+b_5\eta_V+c_5\eta_V^2)\eta_D
\nonumber\\
\end{eqnarray}
with the parameters $a_{4,5}$, $b_{4,5}$ and $c_{4,5}$ given in Table \ref{table3}.

\subsection{Special Points in the M - R diagram}

It can be seen in Fig. \ref{fig5} that varying the Lagrangian parameters $\eta_D$ and $\eta_V$, one can define subsets of EoS which correspond to mass-radius curves that intersect with each other, whereby the crossing points are collimated 
in a narrow range, referred to as SP.
The SP phenomenon was first investigated on the basis of an analysis of the TOV equations for hybrid EoS in Ref. \cite{Yudin:2014mla}, where for the quark matter EoS a constant speed of sound (CSS) form was assumed and a SP was obtained when the bag constant and, thus, the onset of deconfinement was varied. 
Subsequent thorough analyses within the CSS parameterization of the quark matter EoS in Refs. \cite{Cierniak:2020eyh,Blaschke:2020vuy,Cierniak:2021knt,Cierniak:2021vlf} have mapped out the region of possible SP positions in the M - R plane when the two remaining parameters of the CSS EoS, squared speed of sound and reference pressure, were varied. 
SPs (and, thus, hybrid stars) were found up to masses of $\rm 4~M_\odot$. 
This value exceeds the Rhoades-Ruffini bound 
\cite{Rhoades:1974fn} for the upper limit of the maximum mass of NSs. 
This finding also contradicts the common sense that the occurrence of quark matter in NS interiors softens the EoS so that even reaching the well-observed lower limit for maximum masses of $\rm 2~M_\odot$ should pose a problem for hybrid stars.
The physical reason for this is that the deconfinement of quark matter softens the hybrid EoS only at the densities close to the phase transition because of the jump of density at constant pressure.
At the same time, significant stiffness of the quark EoS itself provides stability of very heavy NSs.
Such a stiffening of quark EoS can be observed in Fig. \ref{fig:cs2nB}, which indicates that the quark speed of sound significantly exceeds the hadronic one.

In Ref. \cite{Yudin:2014mla} the existence of an SP and its location were thoroughly derived with the necessary mathematical strictness based on the assumption that the speed of sound of quark matter can be approximated by a constant.
It is remarkable that SPs also appear within the present approach, which is not restricted to a constant speed of sound and asymptotically reaches the conformal limit \cite{Ivanytskyi:2022bjc}.
This gives an explicit example that the existence of SPs is not bound to CSS-type EoSs so analyzing SPs may allow probing parameters of more realistic models of quark matter.
Such an analysis corresponding to the RDF approach is presented below.

The SPs, marked in Fig. \ref{fig5} by blue, magenta, and orange, are obtained for constant values of the vector coupling.
These SPs are obtained by variation of the diquark coupling leading to a family of mass-radius curves.
In Ref. \cite{Shahrbaf:2021cjz} the SPs obtained in this way were argued to be the most important ones for the phenomenology of NSs because of their relation to the NS maximum masses.
For a further discussion of the phenomenology of SPs see Ref. \cite{Shahrbaf:2021cjz}, where the nonlocal NJL model was employed as an EoS for cold and dense color superconducting quark matter. 
One of the phenomenological laws for SPs  conjectured first 
in \cite{Blaschke:2022gql}
relates the maximum mass $\rm M_{\rm max}$ of hybrid stars with the SP mass $\rm M_{\rm SP}$ and the onset mass $\rm M_{\rm onset}$ of the deconfinement transition by
\begin{equation}
 \label{eq:Monset}
 \rm M_{\rm max}=
 M_{\rm SP}+\delta\,|M_{\rm onset}^* - M_{\rm onset}|^\kappa,
\end{equation}
where $\rm M_{\rm onset}^*$, $\kappa\simeq 2$ and the small positive $\delta$ are constant parameters.
As compared to Ref. \cite{Blaschke:2022gql}, where the value of ${\rm M}_{\rm onset}^*$ was only roughly estimated, in the present work we can provide an independent confirmation of the relation (\ref{eq:Monset}) and more precisely quantify its parameters.
We focus on the SPs obtained by varying the diquark coupling at a fixed value of the vector one.
This is sufficient to constrain the values of the vector and diquark couplings of the quark model Lagrangian 
(\ref{I}).

\begin{center}
\begin{table}[t]
\begin{tabular}{ |c|c|c|c|c|c|c|} 
\hline
\multirow{2}{*}{SP}  &
$\rm M_{SP}$ & $\rm R_{SP}$ & $\eta_V$ & $\eta_D$ & $\rm M_{onset} $ & $\rm M_{max}$  \\
 & $\rm[M_{\odot}]$ &[km] & & & $\rm [M_{\odot}]$ & $\rm [M_{\odot}]$  \\
\hline
\multirow{5}{*}{\begin{turn}{90}blue\end{turn}}  &
\multirow{5}{*}{1.973}                           &
\multirow{5}{*}{11.06}                           &
\multirow{5}{*}{0.230}      & 0.749 & 0.251 & 2.044 \\ \cline{5-7}
                      & & & & 0.740 & 0.506 & 2.011 \\ \cline{5-7}
                      & & & & 0.731 & 0.826 & 1.986 \\ \cline{5-7}
                      & & & & 0.721 & 1.169 & 1.974 \\ \cline{5-7}
                      & & & & 0.711 & 1.483 & 1.976 \\ \hline
\multirow{5}{*}{\begin{turn}{90}magenta\end{turn}} &
\multirow{5}{*}{2.092}                           &
\multirow{5}{*}{11.46}                           &
\multirow{5}{*}{0.290}      & 0.760 & 0.251 & 2.159 \\ \cline{5-7}
                      & & & & 0.753 & 0.506 & 2.130 \\ \cline{5-7}
                      & & & & 0.745 & 0.826 & 2.104 \\ \cline{5-7}
                      & & & & 0.737 & 1.169 & 2.094 \\ \cline{5-7} 
                      & & & & 0.730 & 1.483 & 2.095 \\ \hline
\multirow{5}{*}{\begin{turn}{90}orange\end{turn}}   &
\multirow{5}{*}{2.207}                           &
\multirow{5}{*}{11.85}                           &
\multirow{5}{*}{0.35}       & 0.770 & 0.251 & 2.267 \\ \cline{5-7}
                      & & & & 0.764 & 0.506 & 2.241 \\ \cline{5-7}
                      & & & & 0.757 & 0.826 & 2.218 \\ \cline{5-7}
                      & & & & 0.752 & 1.169 & 2.210 \\ \cline{5-7}
                      & & & & 0.747 & 1.483 & 2.209 \\ \hline
\end{tabular}
\caption{Parameters of the SPs and of the mass-radius curves intersecting in those SPs. The columns represent the color of the SPs according to Fig.~\ref{fig5}, their mass $\rm M_{SP}$ and radius $\rm R_{SP}$, as well as the vector $\eta_V$ and diquark $\eta_D$ couplings of the corresponding mass-radius curves along with their onset mass of quark deconfinement $\rm M_{onset}$ and the maximum mass $\rm M_{max}$.}
\label{table4}
\end{table}
\end{center}

The values from Table \ref{table4} evidence that $\rm M_{\rm max}$ is a non-monotonous function of $\rm M_{\rm onset}$ with a shallow minimum at about $1.25~\rm M_\odot$.
The minima of $\rm M_{\rm max}$ representing different SPs are very close to each other. 
The simplest functional dependence to capture these features is a parabolic one with a universal value of the parameter $\rm M_{\rm onset}^*$.
These conclusions are independently confirmed by fitting $\rm M_{\rm SP}$, $\delta$, $\rm M_{\rm onset}^*$ and $\varkappa$ to the data points.
The obtained values of $\rm M_{\rm onset}^*$ vary within less than 1\%, while $\varkappa=1.9-2.1$.
Thus, in what follows we assume $\varkappa=2$ and a universal value of $\rm M_{\rm onset}^*=1.254~M_\odot$ fitted to the data points.
The fitted curves obtained under these assumptions describe very well $\rm M_{\rm max}$ as a function of $\rm M_{\rm onset}$.
The obtained values of $\rm M_{\rm SP}$ are given in Table \ref{table4} and perfectly match the SPs shown in Fig. \ref{fig5}.
These values are precisely described by the linear dependence
\begin{eqnarray}
 \label{eq:M_SP_fit}
{\rm M}_{\rm SP}=k_{{\rm M}_{\rm SP}}\eta_V+b_{{\rm M}_{\rm SP}}
\end{eqnarray}
with $k_{\rm M_{\rm SP}}=1.950~{\rm M}_\odot$ and $b_{\rm M_{\rm SP}}=1.525~{\rm M}_\odot$.
We find that $\delta$ obtained by fitting Eq. (\ref{eq:Monset}) to the values of $\rm M_{max}$ and $\rm M_{onset}$ from Table \ref{table4} is also linear in $\eta_V$, i.e.
\begin{eqnarray}
 \label{eq:delta_fit}
\delta=k_\delta\eta_V+b_\delta\, ,
\end{eqnarray}
where $k_\delta=-0.096~{\rm M}_\odot^{-1}$ and $b_\delta=0.093~{\rm M}_\odot^{-1}$.
Despite the fact that Eqs. (\ref{eq:M_SP_fit}) and (\ref{eq:delta_fit}) are obtained based on fitting three data points, the corresponding values of the vector coupling $\eta_V$ cover a rather wide range.
Extending this range to smaller values does not make physical sense since the corresponding mass-radius curves violate the observational constraints on $\rm M_{\rm max}$, while too large values of $\eta_V$ exclude the possibility of constructing a hybrid EoS by means of the Maxwell construction.

\begin{figure}[t]
\includegraphics[width=\columnwidth]{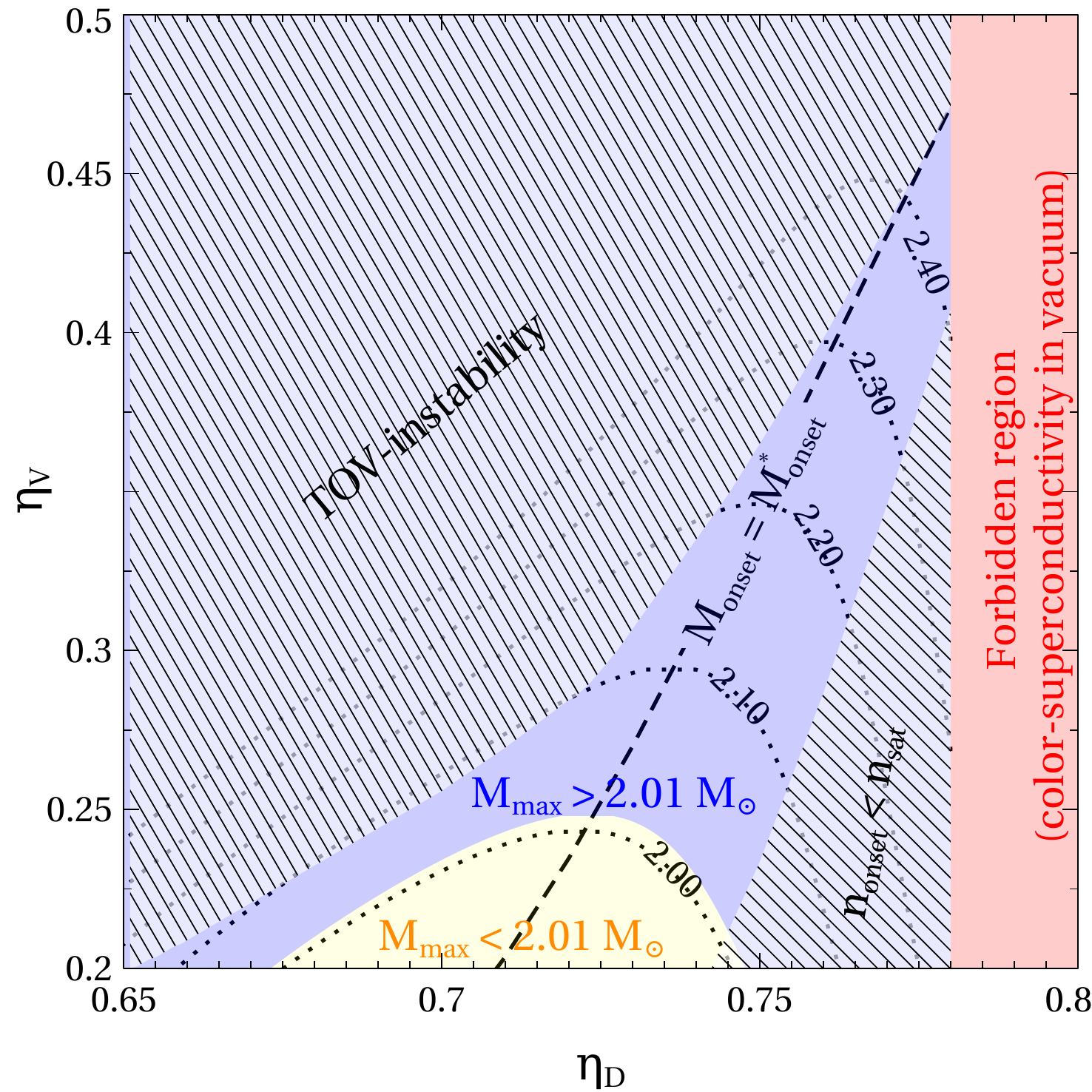} 
\caption{The blue shaded region of the values of vector and diquark couplings is consistent with the observational constraints on the maximum mass of NSs from Ref. \cite{Fonseca:2021wxt}, where $\rm M_{\rm max}\ge2.01~M_\odot$.
Parameter sets in the canary shaded region violate this constraint since for them $\rm M_{\rm max}<2.01~M_\odot$. 
The red region is forbidden since it violates the requirement of the absence of color-superconductivity in the vacuum.
The hatched regions are excluded by the requirements of either having the onset of deconfinement above the saturation density or providing stability of the quark branch of the mass-radius curve obtained as a solution of the TOV equation.
Parameter sets along the dotted curves correspond to the values of $\rm M_{ max}=2.0 (0.1) 2.4~M_\odot$, as indicated in the figure.
For values of $\eta_D$ and $\eta_V$ along the dashed curve holds $\rm M_{\rm onset}=M_{\rm onset}^*$.}
\label{fig7}
\end{figure}

\begin{figure}[t]
\includegraphics[width=\columnwidth]{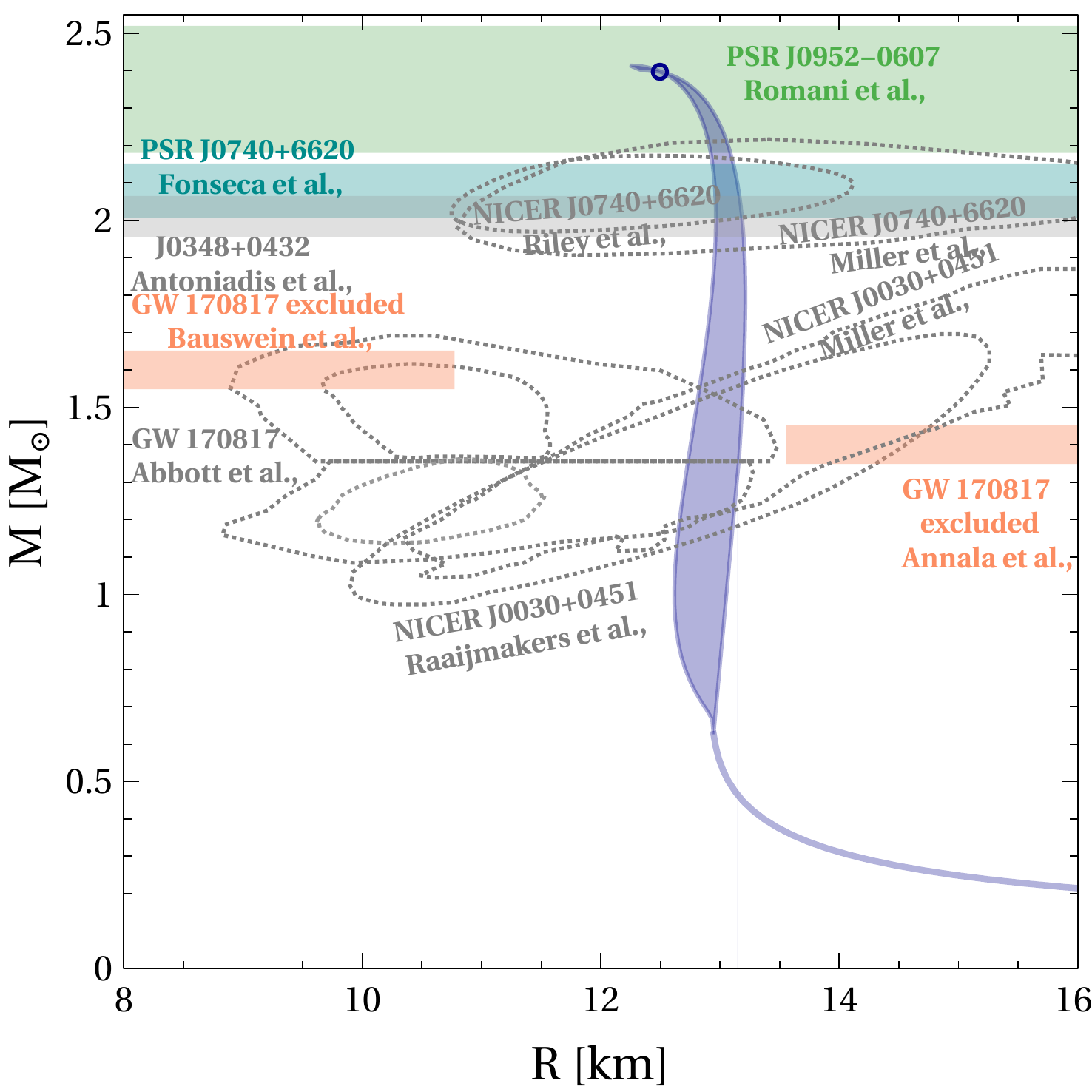} \caption{The same as in Fig. \ref{fig5} but for the value $\eta_V=0.452$ obtained by fitting the $\omega$-meson mass and the range $\eta_D=0.775-0.780$ (see text for details).
The SP is indicated by a filled circle.
With this range of hybrid EoS the black widow pulsar PSR J0952-0607 with a mass of $2.35^{+0.17}_{-0.17}~{\rm M}_\odot$ \cite{Romani:2022jhd} can be described as a hybrid star with a color superconducting quark matter core.}
\label{fig:M_R-etav0452}
\end{figure}

The observational constraint on the NS maximum mass \cite{Fonseca:2021wxt} $\rm M_{\rm max}\ge2.08{\pm 0.07}~M_\odot$ along with Eqs. (\ref{eq:Monset}) - (\ref{eq:delta_fit}) allows us to limit the onset mass of quark deconfinement $\rm M_{\rm onset}$. 
Thus, requiring $\rm M_{\rm max}\ge2.01~M_\odot$ and using Eq. (\ref{eq:Monset_fit}) we can define the region in the $\eta_V-\eta_D$ plane, which is consistent with the above constraint on the NS maximum mass.
Fig. \ref{fig7} demonstrates this region.
At a given value of the vector coupling the onset mass of quark matter is controlled by the diquark one, thus $\rm M_{onset}$ being a decreasing function of $\eta_D$.
At too large values of the onset mass the quark branch of the mass-radius curve is unstable indicated by the negative squared frequency of the fundamental mode of radial oscillations~\cite{1965gtgc.book.....H,1966ApJ...145..505B,1995A&A...294..747G, DiClemente:2020szl}.
We exclude the corresponding region of the $\eta_V-\eta_D$ plane in Fig. \ref{fig7}.
We also exclude the region of small $\rm M_{onset}<0.243~M_\odot$, which corresponds to the quark onset density $n_{\rm onset}$ coinciding with the saturation density of normal nuclear matter $n_{\rm sat}$.
This value is not certain and is chosen as an estimate of the lower limit of $n_{\rm onset}$.
The discussed requirements strongly constrain the values of the vector and diquark coupling of the microscopic Lagrangian assuming them to be in the range either $\eta_V\simeq 0.25-0.47$ and $\eta_D\simeq0.70-0.78$ or $\eta_V\lesssim0.25$ and $\eta_D\lesssim 0.70$. 
We note the similarity in the shape of the allowed region and the range of maximal masses with Fig. 3 of Ref. \cite{Baym:2019iky}.

As is seen from Eq. (\ref{eq:Monset}) the maximum mass of s stellar sequence is never below the mass of the corresponding SP, while its smallest value $\rm M_{\rm max}=M_{\rm SP}$ is obtained at $\rm M_{\rm onset} = M_{\rm onset}^*$.
In other words, for all sequences with  $\rm M_{\rm onset} = M_{\rm onset}^*$, the configuration of the SP coincides with the one of the maximum mass. 
The regions of the $\eta_V-\eta_D$ plane, which are to the left and to the right from the $\rm M_{\rm onset} = M_{\rm onset}^*$ contour shown on Fig. \ref{fig7}, correspond to the quark onset mass above and below $\rm M_{onset}^*$, respectively.

Fig. \ref{fig7} also explains why our analysis excludes small values of the vector coupling.
For example, at $\eta_V=0.2$ the allowed range of the diquark coupling is $\eta_D=0.66-0.675$.
None of the corresponding mass-radius curves is consistent with the observational data from the pulsar PSR J0740+6620 provided by Ref. \cite{Riley:2021pdl}, while agreement with the results of Ref. \cite{Miller:2021qha} is marginal.
Therefore, we discredit $\eta_V\le0.2$.

Furthermore, the physical range of the vector coupling $\eta_V$ can be estimated by fitting the vacuum value of the vector $\omega$-meson mass $M_\omega$.
This $M_\omega$ can be extracted from the pole of the propagator of an auxiliary vector field, which is introduced to the consideration via bosonization of the vector interaction Lagrangian (\ref{II}) by means of the
Hubbard-Stratonovich transformation as it was performed in Ref. \cite{Ivanytskyi:2022oxv}.
Using $M_\omega=782$ MeV \cite{PhysRevD.98.030001} we obtain $\eta_V=0.452$.
This value of the vector coupling along with the constraints presented in Fig. \ref{fig7} allows us to restrict the diquark coupling to the narrow interval $\eta_D=0.775-0.780$.
The lower and upper values from this interval correspond to the deconfinement onset mass and the maximum NS mass $\rm M_{\rm onset}=1.30~M_\odot$, $\rm M_{\rm max}=2.40~M_\odot$ and $\rm M_{\rm onset}=0.68~M_\odot$, $\rm M_{\rm max}=2.41~M_\odot$, respectively.
Fig. \ref{fig:M_R-etav0452} shows the corresponding mass-radius diagram,
which respects all the observational constraints discussed before and includes SP with $\rm M_{\rm SP}=2.40~M_\odot$.
We note that the hybrid EoS corresponding to this favorable range of parameter choice $\eta_V=0.452$ and $\eta_D=0.775-0.780$  
can describe the black widow pulsar PSR J0952-0607 with a mass of $2.35{\pm 0.17}~{\rm M}_\odot$ \cite{Romani:2022jhd} as a hybrid star with a color superconducting quark matter core.

\section{Conclusions and discussions}
\label{concl}

We found a thermodynamically equivalent representation to the confining RDF approach in the asymptotically conformal color superconducting quark matter EoS given by the simple ABPR parameterization.
We have established the connection between the parameters of the two approaches. 
This allows us to avoid numerically demanding calculations in constructing such an EoS and provides the community with an accurate and easy-to-use parameterization of this EoS explicitly given in terms of the vector and diquark couplings of the Lagrangian model.
This enables direct probes of the microscopic parameters of quark matter, i.e. the vector and diquark couplings,
by studying the phenomenology of NSs.

The developed phenomenological EoS was applied to model NSs with quark cores.
We report that SPs in the NS mass-radius diagram are also found in the case of asymptotically conformal quark matter, which is characterized by a significant variation in the speed of sound. 
Variation of the diquark coupling, while the vector one is kept constant, produces a family of hybrid quark-hadron EoSs, which correspond to mass-radius curves intersecting in an SP.
This gives a rule for finding such SPs within microscopic approaches to quark matter.
This allows us to predict the existence of NSs with masses $\rm M_{SP}$ given by Eq. (\ref{eq:M_SP_fit}) if the vector coupling is known precisely enough.

A central result corresponds to constraining the range of possible values of the vector and diquark couplings of the quark model Lagrangian. This was accomplished by using the found empirical relation between the mass of the SP, the maximum mass of the mass-radius curve, and the onset mass for quark deconfinement.
We demonstrated that supplementing this constraint with the value of the vector coupling fitted to the vacuum mass of the $\omega$-meson defines a narrow range of $\eta_V$ and $\eta_D$.
It is remarkable, that this range suggests an early deconfinement of quark matter and deconfinement onset masses below $\rm 1~M_\odot$. 
At the same time, these favored parameter values correspond to high maximum masses and thus are consistent with the existence of heavy NSs with masses up to $\rm 2.4~M_\odot$, which suggests a possible explanation of the nature of the black widow
pulsar PSR J0952-0607 as a hybrid star with a color superconducting quark matter core.

\vspace{0.5cm}
\section{acknowledgments}
C.G. acknowledges the Funda\c c\~ao para a Ci\^encia e Tecnologia (FCT), Portugal, through the IDPASC PT-CERN program and support of the Center for Astrophysics and Gravitation (CENTRA/IST/ULisboa) through grant project No. UIDB/00099/2020 and grant No. PTDC/FIS-AST/28920/2017. The work of O.I. was supported by the program Excellence Initiative--Research University of the University of Wrocław of the Ministry of Education and Science. V.S. acknowledges the support from the Funda\c c\~ao para a Ci\^encia e Tecnologia (FCT) within the projects EXPL/FIS-AST/0735/2021, UIDB/04564/2020, UIDP/04564/2020. D.B. acknowledges the support from the Polish National Science Center under grant No. 2019/33/B/ST9/03059. 

\bibliography{references}

\end{document}